\documentclass[twocolumn]{aastex631}

\usepackage{amsmath}
\DeclareUnicodeCharacter{2212}{-}
\graphicspath{{./}{figures/}}

\shorttitle{Machine learning classification of WR stars}
\shortauthors{Kar \& Bhattacharya et al.}
\begin{document}

\title{Classification of Wolf Rayet stars using Ensemble-based Machine Learning algorithms}

\author[0000-0001-7874-0218]{Subhajit Kar} \altaffiliation{ Author contributed equally} 
\affiliation{S.N. Bose National Centre for Basic Sciences,
Kolkata, India} 

\correspondingauthor{Subhajit Kar}
\email{subhajit0596@gmail.com}

\author[0009-0007-8229-3036]{Rajorshi Bhattacharya} \altaffiliation{ Author contributed equally}
\affiliation{Department of Physics and Astronomy, University of New Mexico, 87131 Albuquerque, NM, USA}

\correspondingauthor{Rajorshi Bhattacharya}
\email{rbhattacharya1995@unm.edu}

\author[0000-0002-5440-7186]{Ramkrishna Das}
\affiliation{S.N. Bose National Centre for Basic Sciences,
Kolkata, India}

\author[0000-0003-0615-1785]{Ylva Pihlström} 
\affiliation{Department of Physics and Astronomy, University of New Mexico, 87131 Albuquerque, NM, USA}
\affiliation{National Radio Astronomy Observatory, Pete V. Domenici Science Operations Center, Socorro, NM 87801, USA}

\author[0000-0002-8069-8060]{Megan O.\ Lewis}
\affiliation{Leiden Observatory, Leiden University, PO Box 9513, 2300 RA Leiden,
The Netherlands}

%% Note that the \and command from previous versions of AASTeX is now
%% depreciated in this version as it is no longer necessary. AASTeX 
%% automatically takes care of all commas and "and"s between authors names.

%% AASTeX 6.2 has the new \collaboration and \nocollaboration commands to
%% provide the collaboration status of a group of authors. These commands 
%% can be used either before or after the list of corresponding authors. The
%% argument for \collaboration is the collaboration identifier. Authors are
%% encouraged to surround collaboration identifiers with ()s. The 
%% \nocollaboration command takes no argument and exists to indicate that
%% the nearby authors are not part of surrounding collaborations.

%% Mark off the abstract in the ``abstract'' environment. 
\begin{abstract}
We develop a robust Machine Learning classifier model utilizing the eXtreme-Gradient Boosting (XGB) algorithm for improved classification of Galactic Wolf-Rayet (WR) stars based on Infrared (IR) colors and positional attributes. For our study, we choose an extensive dataset of 6555 stellar objects (from 2MASS and AllWISE data releases) lying in the Milky Way (MW) with available photometric magnitudes of different types including WR stars. Our XGB classifier model can accurately (with an 86\% detection rate) identify a sufficient number of WR stars against a large sample of non-WR sources. The XGB model outperforms other ensemble classifier models such as the Random Forest. Also, using the XGB algorithm, we develop a WR sub-type classifier model that can differentiate the WR subtypes from the non-WR sources with a high model accuracy ($>60\%$). Further, we apply both XGB-based models to a selection of 6457 stellar objects with unknown object types, detecting 58 new WR star candidates and predicting sub-types for 10 of them. The identified WR sources are mainly located in the Local spiral arm of the MW and mostly lie in the solar neighborhood.

\end{abstract}

%% Keywords should appear after the \end{abstract} command. 
%% See the online documentation for the full list of available subject
%% keywords and the rules for their use.
%\keywords{Random Forests (1935), Two-color diagrams (1724), Wolf-Rayet stars (1806), Stellar classification (1589), Massive stars (732), Infrared photometry (792)}

%% From the front matter, we move on to the body of the paper.
%% Sections are demarcated by \section and \subsection, respectively.
%% Observe the use of the LaTeX \label
%% command after the \subsection to give a symbolic KEY to the
%% subsection for cross-referencing in a \ref command.
%% You can use LaTeX's \ref and \label commands to keep track of
%% cross-references to sections, equations, Tables, and figures.
%% That way, if you change the order of any elements, LaTeX will
%% automatically renumber them.
%%
%% We recommend that authors also use the natbib \citep
%% and \citet commands to identify citations.  The citations are
%% tied to the reference list via symbolic KEYs. The KEY corresponds
%% to the KEY in the \bibitem in the reference list below. 

\section{Introduction} \label{sec:intro}
Wolf-Rayet (WR) stars belong to the evolved class of Population I stellar objects, mainly arising from the evolutionary trajectory of massive O-type Main Sequence (MS) stars. In a binary system, MS O-type objects ($\geq15\,M_{\odot}$) undergo mass transfer to produce WR stars, either via Roche Lobe overflow or through the formation of a common envelope; while single WR star ($\geq25\,M_{\odot}$) formation is governed by the interplay of mass-loss mechanisms and rotational mixing processes. Spectroscopically, based on elemental abundances WR stars can be classified as: WN (He and N), WC (He and C), and WO (C and O) types. Based on the relative line strengths of the diagnostic emission lines \citep{2001NewAR..45..135V}, these stars are further classified: WR-Late (WRL; with lower-ionization lines due to low temperature) and WR-Early (WRE; hotter subtypes emitting lines from higher ionization states). Some WNL-type stars show weak emission lines of hydrogen present in their outer envelopes. Such objects are called non-classical WR stars and are suffixed with 'h'. With surface temperature ranging from $\mathrm{3\times10^{4}-2\times10^{5}}$\,K and luminosity spanning from $10^{4}-10^{6}\,L_{\odot}$, WR stars are known for broad emission lines propelled by supersonic stellar winds ($v_{\infty}$=1000$-$3000\,$kms^{-1}$) which likely undergo Type Ib/Ic supernovae explosion, thereby contributing significantly towards both mechanical and chemical enrichment of the surrounding interstellar medium and providing the necessary habitat for young star formation. WR stars often occur in clusters, making them suitable\,tracers of massive star formation regions in the MW \citep{2012MNRAS.419.1860D} and in Starburst Galaxies \citep{Schaerer_1998}. WR stars exhibit strong free-free emission leading to excess flux in the near-IR (NIR) bands making them distinguishable from other stellar objects. 

From IR \citep{2009PASP..121..591M, 2012AJ....143..149S, 2015MNRAS.447.2322R} and optical surveys \citep{10.1093/mnras/stz3614}, a significant population\,($\sim$670, \citet{2015MNRAS.447.2322R}) of WR stars in the MW has been detected, while simple population models \citep{2015MNRAS.447.2322R} predict that around 2000 WR stars may be present. However, distinguishing WR stars from other stellar objects solely based on optical colors and magnitudes has been challenging because of strong interstellar dust extinction. Meanwhile, the IR bands suffer from much less obscuration than the optical bands, thus benefiting the detection of WR stellar candidates. With the advent of modern instruments such as JWST, Roman telescope, etc., we are bound to discover several thousands of such evolved stellar objects lying in the Local Group of galaxies. 

Selection criteria based on broad-band IR (2MASS and GLIMPSE) colors and magnitudes by \citet{2007MNRAS.376..248H, 2011AJ....142...40M} have been successful in filtering non-WR from WR candidates. Free-free emission and/or emission from circumstellar dust leads to color-excess in the IR which benefits WR candidate identification based on the color-color diagrams. With the increase in IR survey data over the last few decades (2MASS, IRTF, WISE, Herschel, etc.) it becomes cumbersome to manually perform this task. Machine learning (ML) techniques are well suited for handling large datasets and offer a powerful and flexible approach to stellar classification. Previously, distance-based algorithms such as K-Nearest Neighbor (KNN) and Support Vector Machine (SVM) have been used for classification of WR stars but on a very small dataset of objects \citep{2018MNRAS.473.2565M, 2021ApJ...913...32D}. A comparative study by \citet{Yoshino2023ExploringXA} showed that ensemble-based algorithms such as XGBoost (XGB) and Random Forest (RF) perform consistently better in stellar classification than a simple Decision tree, KNN, or SVM models.

In our current study, we use the XGB methods to build the most efficient stellar classifier customized to proficiently identify WR stars based on the IR colors and positional coordinates of the objects. The method is designed to identify WR stars in large datasets comprised of various types of IR-bright objects. We describe the methodology used to develop the classifier 
models in Sec.\,\ref{sec:methodology}. Our results are given in Sec.\,\ref{sec:results} including a comparative evaluation with other ensemble methods such as Random Forest. Also, we conduct novel research on the application of ML models to distinguish the WR subtypes (WC and WN) from the non-WR objects (in Sec.\,\ref{subsec:wr_subtype}). A list of new WR stars (and their chemical subtypes) predicted by both models on an unlabelled and unseen Galactic stellar dataset is presented in Sec. \ref{subsec:new_wr}. Further, we discuss the performances of both the models against color-selection methods in Sec.\,\ref{sec:discussion}. Concluding remarks are presented in Sec.\,\ref{sec:conclusion}.
%%%%%%%%%%%%%%%%%%%%%%%%%%%%%%%%%%%%%%%%%%%%%%%%%%%%%%%%%%%%%%%%%%%%%%%%%%%%%%%%%%%%%%
\section{Data Sources}
For this study, we utilize the apparent photometric magnitudes of stellar objects in the IR bands. The NIR and mid-IR (MIR) data used in this study were observed using the 2-Micron All Sky Survey (2MASS) and the Wide-field Infrared Survey Explorer (WISE). The 2MASS photometric data were observed across 3 broad-bands: J (1.25\,$\mu m$), H (1.65\,$\mu m$) and $K_{s}$ (2.16\,$\mu m$) bands while the WISE data were observed in 4 different broad-bands: W1 (3.4\,$\mu m$), W2 (4.5\,$\mu m$), W3 (12\,$\mu m$) and W4 (22\,$\mu m$). We selected all the sources with an available object-type flag from the SIMBAD database and then cross-matched them (within a radius of 5") with the Vizier catalogs of 2MASS and WISE surveys: II/246/out \citep{vizierII246} and II/328/allwise \citep{2014yCat.2328....0C} respectively. To address the dense source distribution near the Galactic Center, we eliminated redundant sources by further cross-matching such sources with a 1” radius. In the dataset, other than WR stars, we consider various types of stellar objects such as MS, Asymptotic Giant Branch (AGB; including the subtypes Mira and OH/IR), Be, Red Super Giant (RSG), Long Period Variable (LPV), Young Stellar Objects (YSO), C, S, High Mass X-ray Binary (HMXB), Emission line (Emline), Yellow Super Giant (YSG), RRLyrae, Ae, Red Giant Branch (RGB), High Proper Motion Star (High PM), RCrB, beta Cepheid Variable (bCepV), Eclipsing Binary (EB), RV Tauri Variable (RVTauV), Orion Variable (OrionV), Classical Cephieds (Ce), Hot Sub-dwarf, Horizontal Branch (HB), Variable, Symbiotic, Pulsar, Blue Super Giant (BSG), post-AGB (pA) and Planetary Nebulae (PNe). 
%%%%%%%%%%%%%%%%%%%%%%%%%%%%%%%%%%%%%%%%%%%%%%%%%%%
\section{Methods}\label{sec:methodology}
In this study, we follow the procedure as described in the following subsections (Sec. \ref{subsec:data}-\ref{subsec:hyper_parameter}). Before defining the ML model, it is crucial to pre-process the dataset which includes sorting the data and selecting the model features.

\begin{figure*}
\gridline{\fig{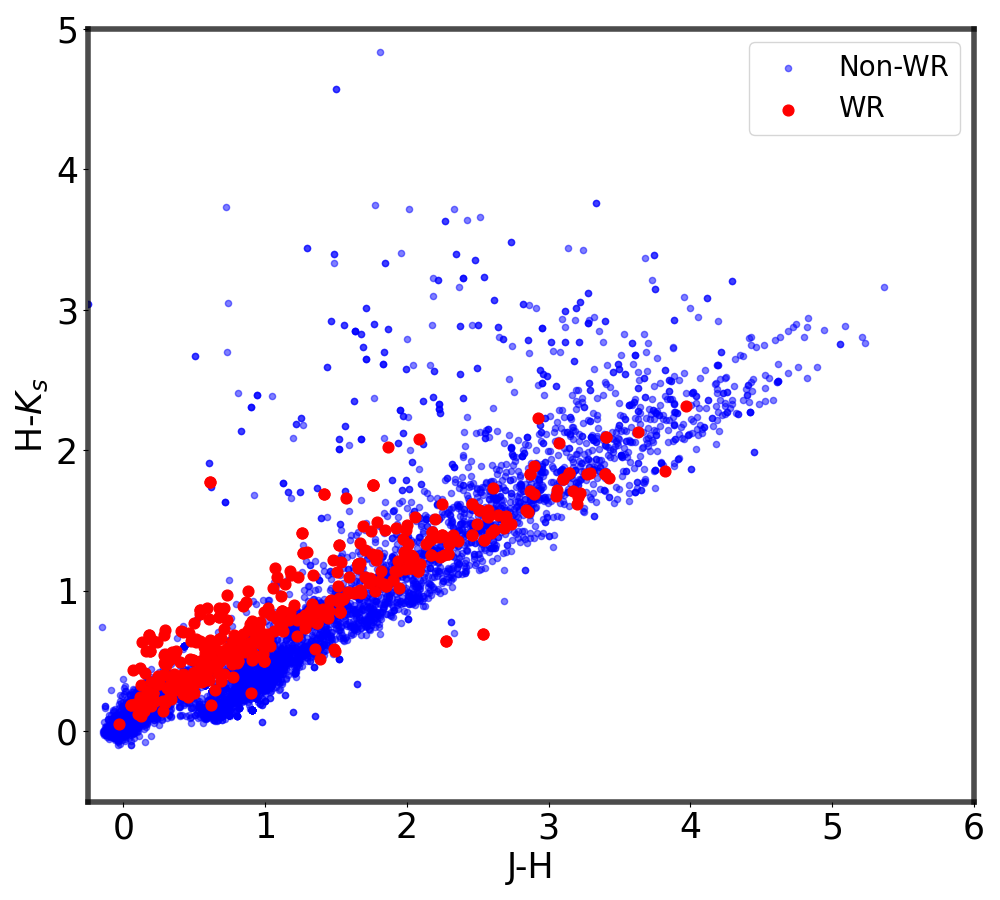}{0.5\textwidth}{(a)}
          \fig{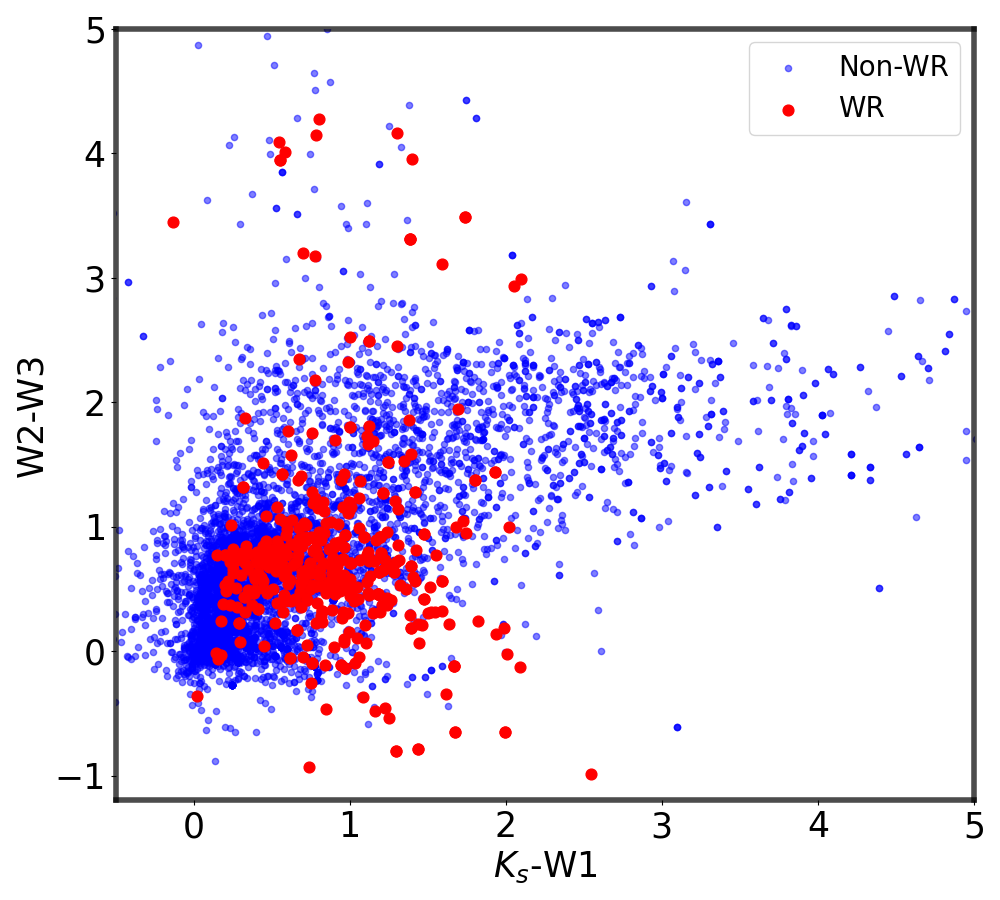}{0.5\textwidth}{(b)}
         }
\gridline{\fig{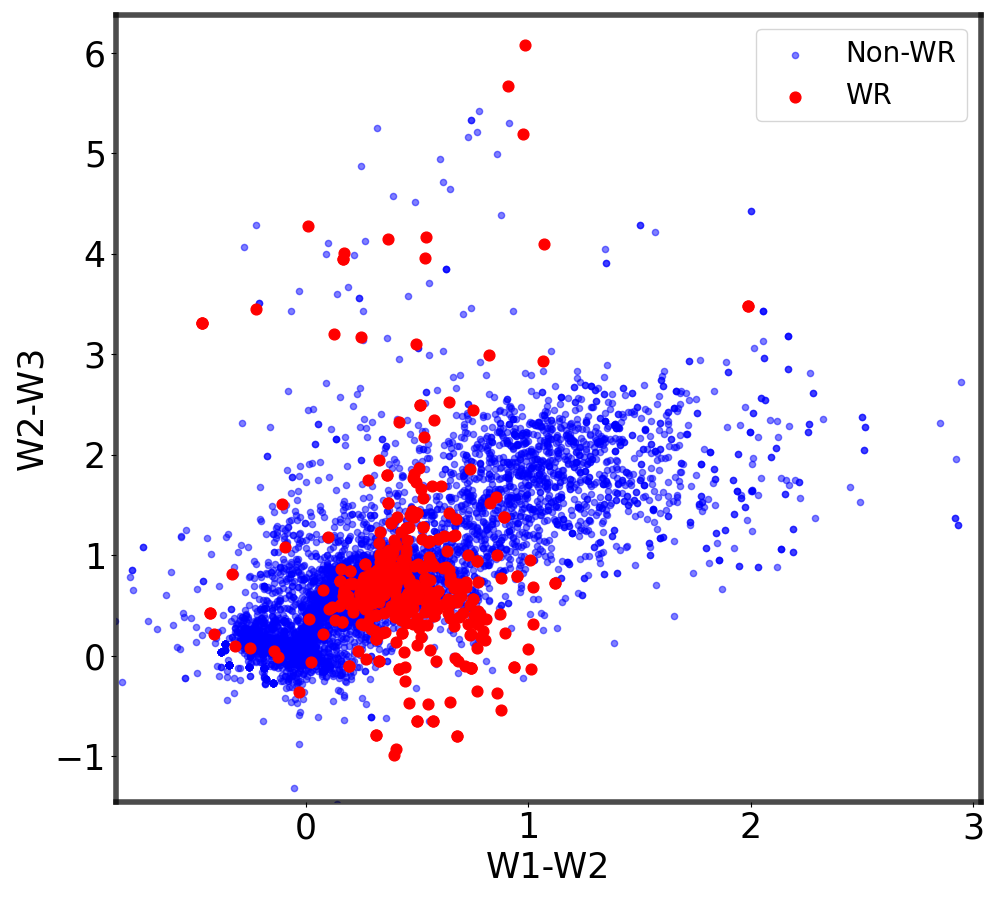}{0.5\textwidth}{(c)}
          \fig{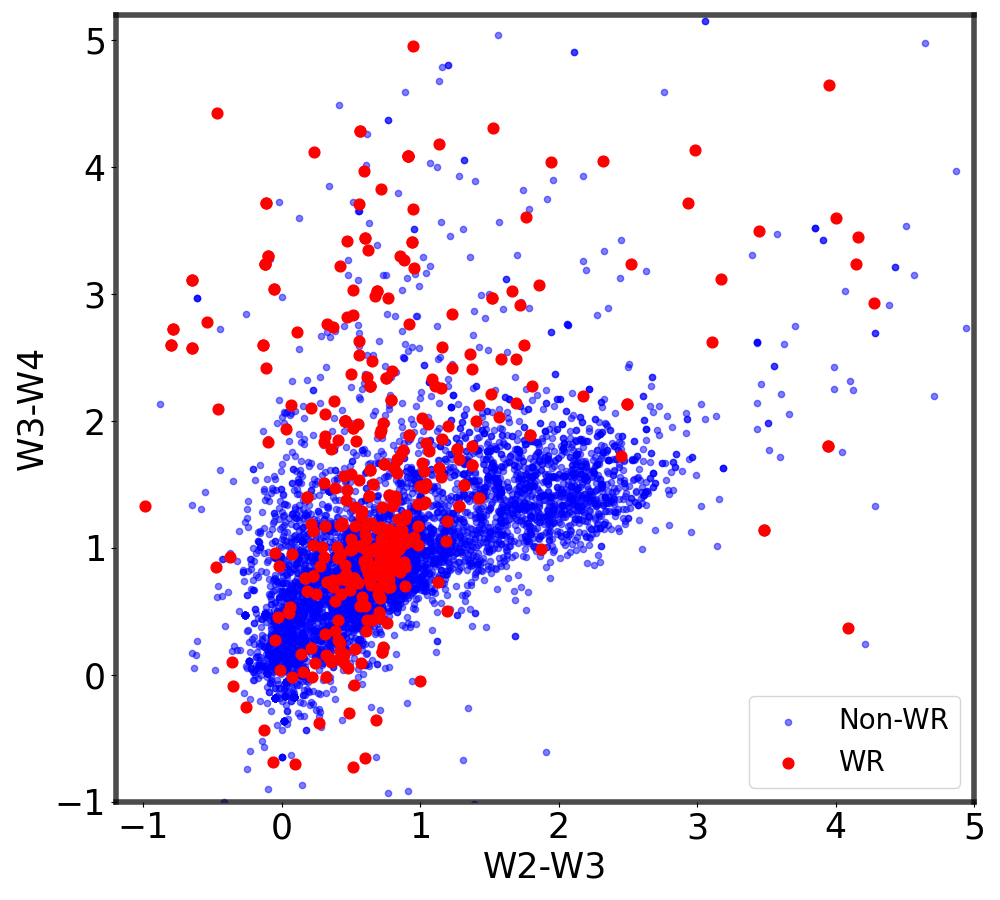}{0.5\textwidth}{(d)}
          }
\caption{IR color-color diagrams depicting distinctive nature of WR (in \textit{red}) and non-WR candidates (in \textit{blue}) in the MW. The non-WR sources include different types of objects such as MS, AGB, Be, RSG, HMXB, LPV etc.} \label{fig:color-color}
\end{figure*}

\subsection{Sorting Data and Features}\label{subsec:data}
We chose stellar sources with Dec $>$ -63.5$^{\circ}$ in order to exclude objects present in the (Large and Small) Magellanic clouds. Thereafter, we cleaned the dataset by choosing 10242 Galactic stellar objects with available apparent magnitudes in both 2MASS ($J$, $H$, $K_{s}$) and WISE ($W1$, $W2$, $W3$, $W4$) bands. Further, we selected objects based on the quality (flagged as A, B, or C ; where flags are defined from A-D, with A as best data with high Signal-to-Noise (S/N) while D as poor without any S/N information) of their 2MASS photometric data (see \citet{2006AJ....131.1163S} for S/N thresholds of the quality flags).

\begin{figure*}
    \centering \includegraphics[width=0.8\textwidth]{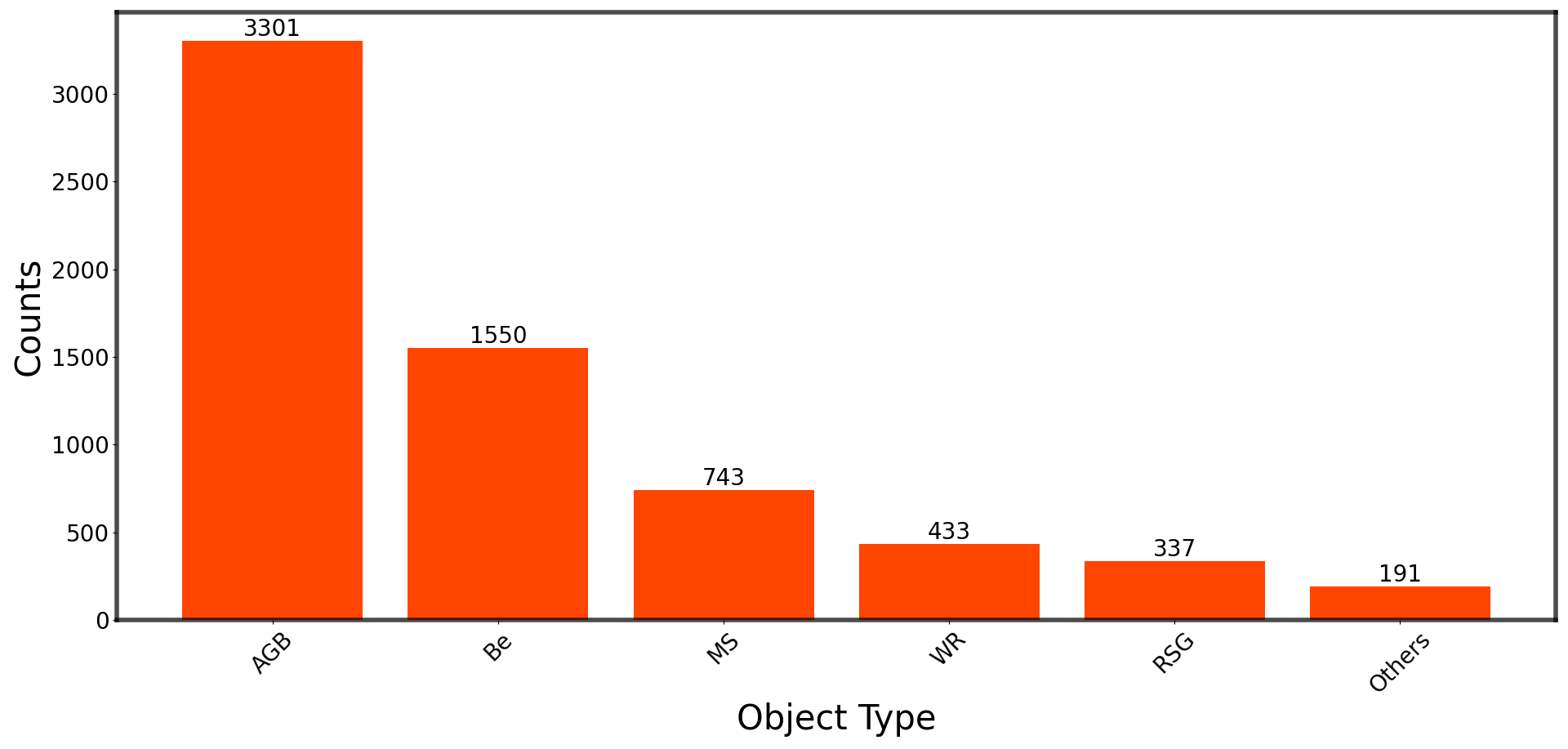}
    \caption{Census of stellar candidates based on object types in Dataset-1.}
    \label{fig:data_sources}
\end{figure*}
\begin{deluxetable}{cc}
\tabletypesize{\scriptsize}
\tablewidth{0pt} 
%\tablenum{1}
\tablecaption{Stellar sources (with population less than 40) categorised as "Others" (see Sec. \ref{subsec:data}).\label{tab:data_sources}}
\tablehead{
\colhead{Object Type} & \colhead{Population} 
} 
%\decimalcolnumbers
\startdata 
HMXB & 36 \\
LPV & 33 \\
C & 31 \\
S & 25 \\
Ae & 13 \\
bCepV & 13 \\
RGB & 8 \\
Star & 8 \\
Emline & 4 \\
BSG &  5 \\
PNe & 3 \\
HB & 2 \\
RVTauV & 2 \\
OrionV & 2 \\
Ce & 2 \\
Symbiotic & 1 \\
Pulsar & 1 \\
Variable & 1 \\ \hline
Total & 191 \\
\enddata
%\tablecomments{}
\end{deluxetable}
\begin{figure*}
    \centering \includegraphics[width=0.5\textwidth]{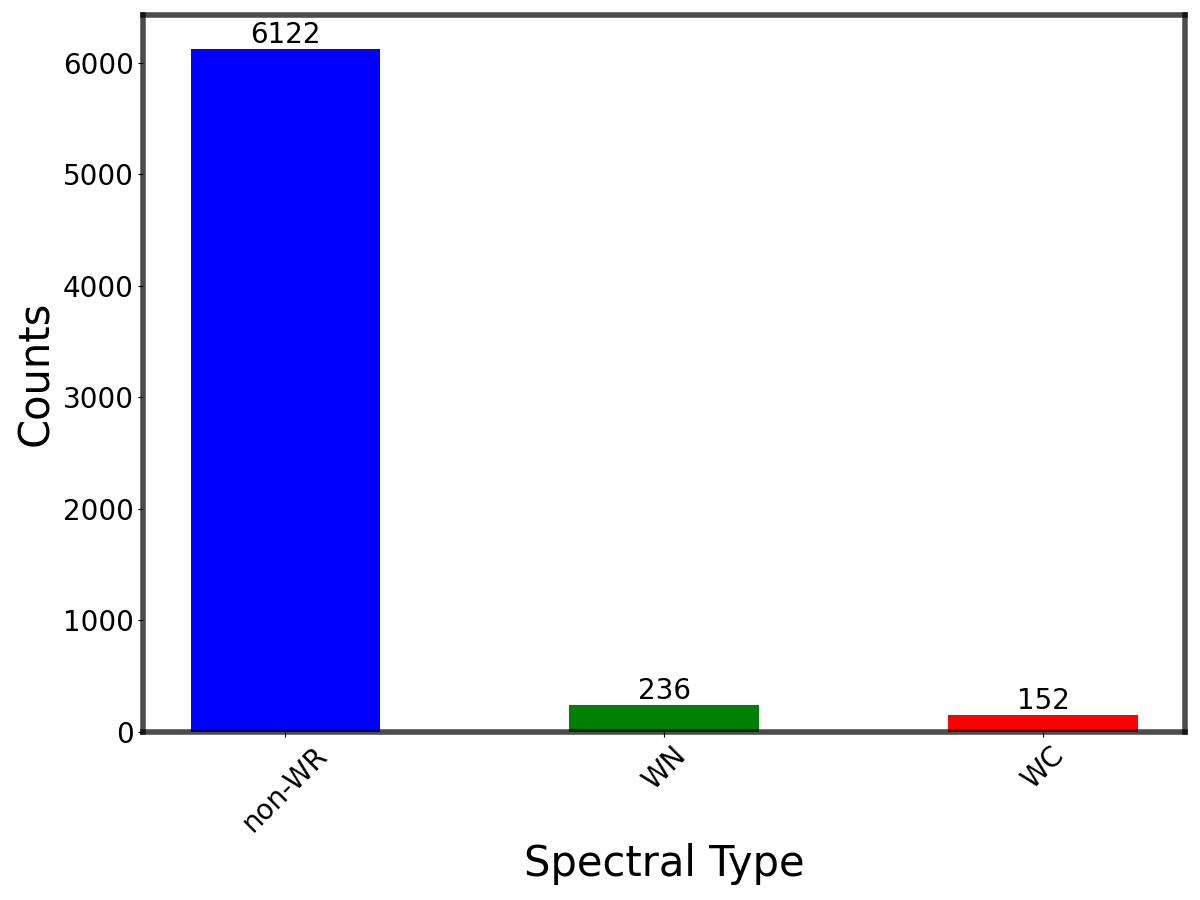}
    \caption{Census of stellar candidates with WR-subtypes in Dataset-2.}
    \label{fig:spec_data_sources}
\end{figure*}

Following \citet{2011AJ....142...40M}, by visual inspection we identify the color space based on the concentration of WR stars in our dataset. Accordingly, we employ the following selection criteria: J $\in$\,$\left[5,17\right]$, H $\in$\,$\left[5,15\right]$, $K_{s}$ $\in$\,$\left[2,14\right]$, W1 $\in$\,$\left[3,15\right]$, W2 $\in$\,$\left[3,15\right]$, W3 $\in$\,$\left[0,10\right]$, and W4 $\in$\,$\left[0,9\right]$. The constrained color space leads to better sampling and increases the chances of WR identification \citep{2011AJ....142...40M}. The color-color plots of the stellar objects in our final dataset are shown in Fig.\,\ref{fig:color-color}. Fig.\,\ref{fig:data_sources} shows the population distribution for different types of objects present in the final dataset (hereafter, Dataset-1 comprising 6555 objects). The majority of the objects in our dataset are AGBs and Be-type stars with WR stars being about 7\% of the total stellar population. Objects with a population lower than 40 are categorized as "Others" (see Table\,\ref{tab:data_sources}). It must be noted that there is a lack of PNe, pAs, and YSOs in our dataset which is mainly due to the color-selection criteria adopted in this study. However, adding these sources would not impact our model, as long as the samples on which we run the ML models on have the same color cuts applied as the training sets. Thereafter, we declare a label column that contains discrete values, i.e. either 1 (for WR) or 0 (for non-WR) as reported in SIMBAD. In our study, the objects are not refined into temperature classes which enables us to introduce a high imbalance in our dataset by treating all non-WR sources (i.e. 6122 objects) as one large (majority) class while the WR stars (i.e. 433 objects) are labeled as the other (minority) class. This makes our (binary classifier) models much easier to implement than earlier (multi-class classification) studies \citep{2018MNRAS.473.2565M}. To treat the highly imbalanced dataset where WR stars belong to the minority class, we use \textit{RandomOversampler}$\footnote{\url{https://imbalanced-learn.org/stable/references/generated/imblearn.over_sampling.RandomOverSampler.html}}$ from \textit{imbalanced-learn}$\footnote{\url{https://imbalanced-learn.org/stable/index.html}}$. Random oversampling is a technique used in ML to address class imbalance in datasets. Class imbalance occurs when one class (the minority class) is significantly underrepresented compared to another class (the majority class). This can lead to biased models that favor the majority class (in this case non-WR sources) and perform poorly on the minority class (WR sources). In random oversampling, examples from the minority class are randomly selected and duplicated until the class distribution is more balanced. By increasing the number of instances in the minority class, the classifier can learn from more representative data and make better predictions for that class. Alternatively, we could have under-sampled the majority class randomly many times, but chose to oversample instead because under-sampling might eventually lead to underfitting in the model. 

For the second part of our study, we utilize a different dataset (Dataset-2) with a separate column in which we further designate the WR subtypes (WN and WC) as two separate entities: non-WR (as 0), WC (as 1) and WN (as 2). As the smaller number of WO-type sources creates a large imbalance that affects the accuracy, we exclude them from the dataset. We further cross-match with SIMBAD to choose WR candidates with known sub-types lying in the same IR color space as the objects in Dataset-1. The population density of different spectral types present in Dataset-2 (with total candidates of 6510) is shown in Fig.\,\ref{fig:spec_data_sources}. In this case, as well, we treat the minority classes (WC and WN) using the random-oversampling method as mentioned earlier. 

We randomly divide both datasets (1 and 2) into two parts: Training datasets (Trd-1 and Trd-2) containing 80\% of their respective samples and testing data sets (Tsd-1 and Tsd-2) containing 20\%. \\
\begin{figure*}
\gridline{\fig{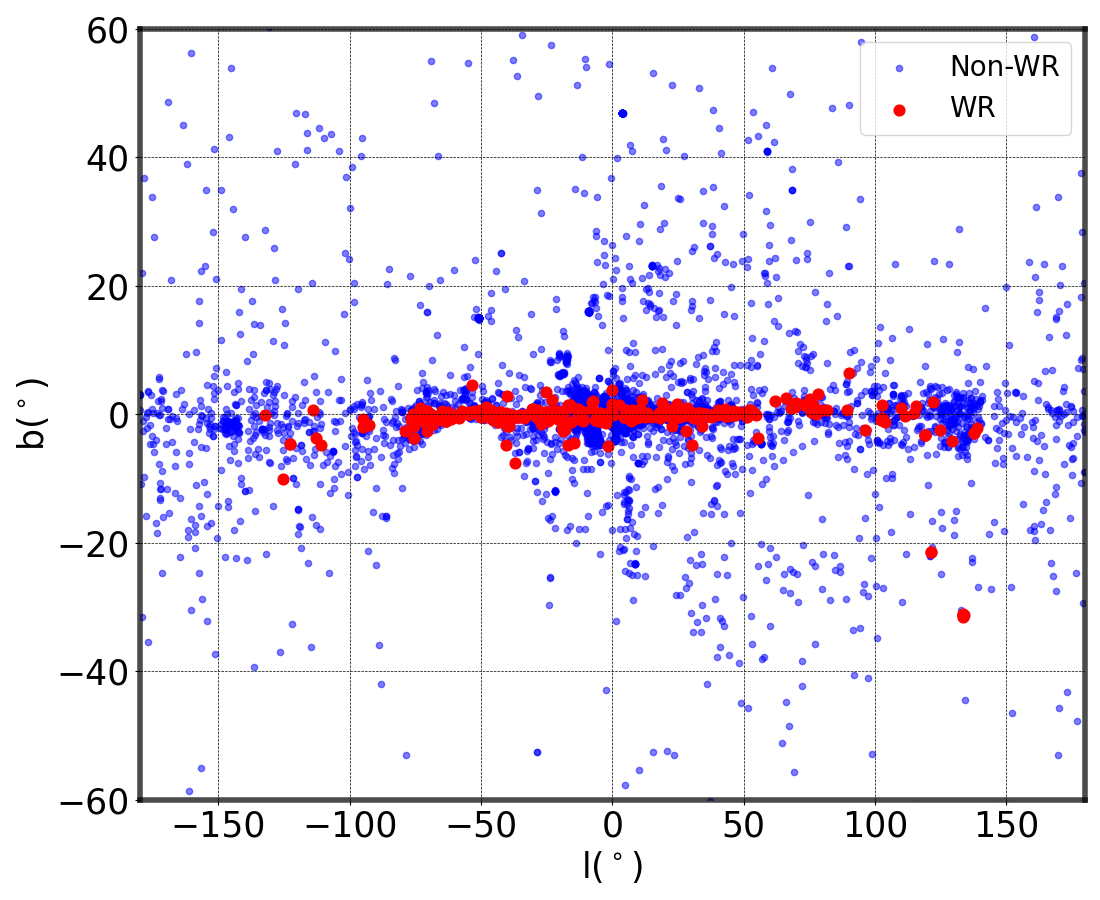}{0.45\textwidth}{(a)}
          \fig{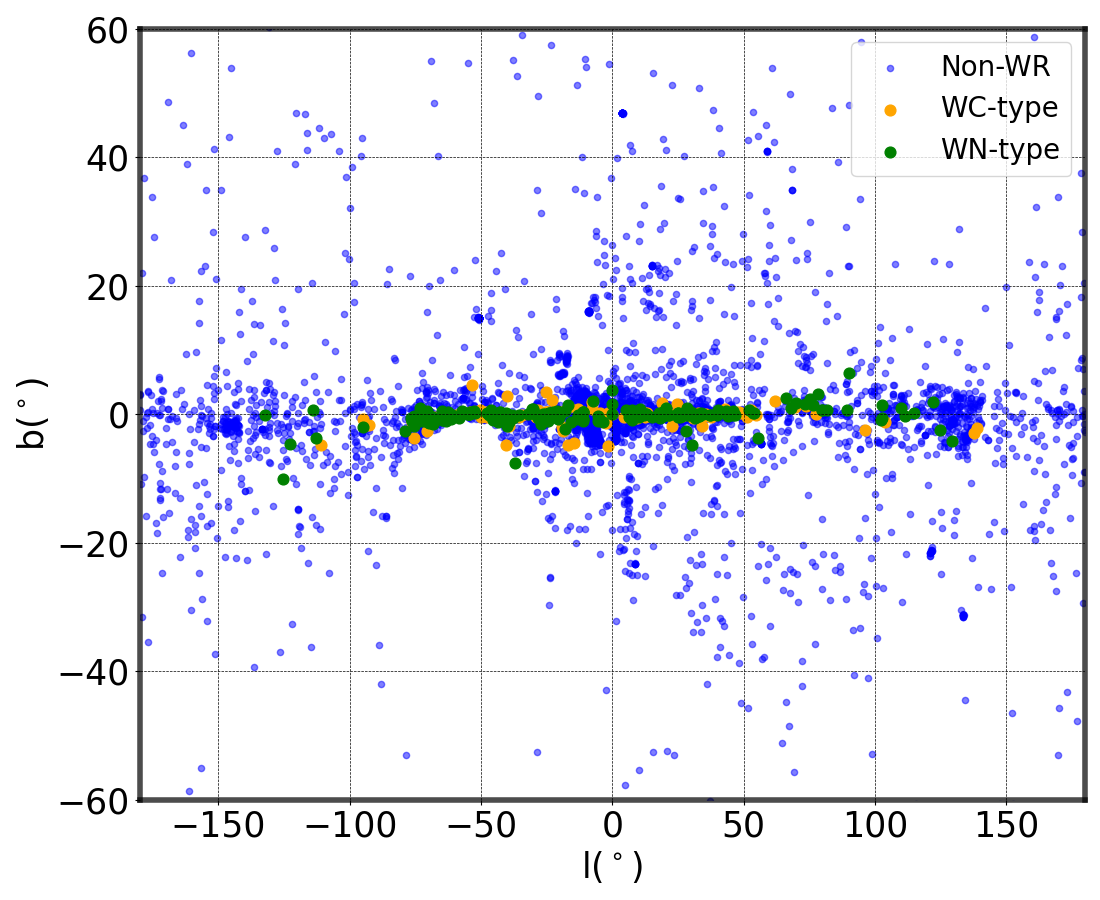}{0.45\textwidth}{(b)}
         }
    \caption{Positional distribution of Galactic stellar sources included in (a) Dataset-1 (\textit{Blue} represents non-WR and \textit{red} represents WR candidates), and (b) Dataset-2 (\textit{Blue} represents non-WR, \textit{orange} represents WC and \textit{green} represents WN-type objects).}
    \label{fig:l_vs_b}
\end{figure*}
As for the model features, we choose a list of features that consists of positional attributes (Right Ascension and Declination) as well as the IR colors ($J-H$, $H-K_{s}$, $K_{s}-W1$, $W1-W2$, $W2-W3$ and $W3-W4$). For this work, we do not include the magnitudes, as neither distance estimates nor extinction values are known for the whole sample. Although IR colors can be affected by distance-dependent extinction, this effect is much smaller than the effect on magnitudes or even optical colors. \citet{10.1093/mnras/stac2760} used the positional coordinates (RA and DEC) along with other observational attributes as model features for the identification of CVs present either in MW or other nearby galaxies. Stars of any particular class are found mostly at certain regions of any galaxy (including the MW) which helps in their identification regardless of the galaxy in which they are found. Therefore, including the positional attributes as model features will surely facilitate the detection of WR stars, especially within the MW. Also, we have a large number of sources spread across the observable disc of the MW without any significant bias in the Galactic plane (see Fig.\,\ref{fig:l_vs_b}). Therefore, our model is well-suited to WR identification across the MW, but we do not recommend its use outside of the MW.

\subsection{Machine Learning models}\label{subsec:machine_learning_models}

We deploy two supervised ML model algorithms: \textit{Gradient Boosting} (GB) \citep{10.1214/aos/1013203451} and \textit{Random Forest} (RF)$\footnote{\url{https://scikit-learn.org/stable/modules/ensemble.html}}$ algorithms provided by \textit{Scikit-learn} \citep{scikit-learn} respectively. Supervised ML models utilize labeled datasets to train algorithms for predictions or decisions for classification problems over very large datasets. By providing the algorithm with labeled examples, supervised learning enables the model to generalize patterns and make accurate predictions on unseen or blind data. Any ensemble classifier is commonly defined using a set of parameters that can comprise of: \textit{n\_estimator} (number of base learners), \textit{max\_features} (maximum number of features for a learner), \textit{max\_depth} (maximum number of nodes in a learner), and \textit{subsample} (random sampling ratio of data to avoid overfitting).   

\subsubsection{XGB Classifier}\label{subsec: xgboost_model}
The XGB$\footnote{\url{https://github.com/dmlc/xgboost}}$\citep{2016arXiv160302754C} library is an advanced implementation of GB classification methods. This boosting technique combines multiple weak learners (\textit{decision trees}) to generate a stronger model. In brief, a \textit{decision tree} breaks down the feature space into smaller sections through repeated splitting based on feature values, creating a tree-shaped structure where each \textit{internal node} represents a decision point, each \textit{branch} signifies the result of that decision, and each \textit{leaf node} represents the final predictions of the model. The predictions from the ensemble of such weak learners are used to estimate the \textit{training loss} function that is used as an input for the next learner. Each learner is constructed sequentially, aiming to minimize the overall loss of the ensemble while incorporating \textit{regularization} to prevent over-fitting. The XGB is faster and more efficient than the standard GB as it can build base learners in parallel for a large dataset. The XGB-tree classifier model not only depends on the common parameters (see Sec.\,\ref{subsec:machine_learning_models}) but also on the \textit{learning\_rate} (step size of learner that modifies the feature weights for the next learner), \textit{min\_child\_weight} (minimum weight for the sum of the features in the learner branch), and \textit{min\_split\_loss} (minimum loss reduction for splitting a leaf node). Earlier studies for example, \citet{2023arXiv230502407G} and \citet{CHAO2019539} showed the efficient capability of XGB algorithms as applied to a regression model to predict rotation periods, and a classifier model to distinguish between stars and galaxies, respectively.

\subsubsection{Random Forest Classifier}\label{subsec: rf_model}
For comparison, we also develop an RF classifier model \citep{2001MachL..45....5B}, which trains the model with several random subsets of the features using an ensemble of \textit{Decision Tree} models. Each tree in the forest is trained on a random subset of the data and a random subset of features, promoting diversity among the trees. The feature selection by the model is done based on the choice of the \textit{entropy} function. The bootstrap aggregation used for estimating the final model prediction (from several trees) reduces over-fitting. RF models work better than a single \textit{decision tree} model because of this bootstrapping. Like any other ensemble model, an RF classifier is mostly based on the choice of the general parameters (as mentioned in Sec.\,\ref{subsec:machine_learning_models}). It is considered that RF classifiers with built-in parallelization are less prone to over-fitting than a sequential GB classifier. Therefore, using an RF model as a benchmark is an effective way to evaluate the performance of the XGB model.
%We follow the same procedure (as mentioned in Sec.\,\ref{subsec: xgboost_model}) to develop the best RF classifier model. 
%The \textit{Bayesian optimization} over a large parameter space ($n\_estimator\,\in$\,[5, 50], $max\_feature$='sqrt', $max\_depth\,\in$\,[1, 50], $min\_samples\_split\,\in$\,[2, 10]) by training and cross-validating the classifier over 5-fold random subsets of the TP. 
%The hyper-tuned model (see Table\,\ref{tab:model_params}) is applied to the TsD and the model metrics are estimated (as shown in Table\ref{tab:metrics}).

\subsection{Model Performance}\label{subsec:model_performance}

The models are evaluated from the confusion matrices that represent the classification of objects present in a test dataset. The model performance is further judged based on the the four important metrics: Accuracy(A), Precision (P), Recall (R), and f1-score (f1):
\begin{equation}\label{accuracy}
    A=\frac{TP+TN}{TP+FP+TN+FN}
\end{equation}
\begin{equation}\label{precision}
    P=\frac{TP}{TP+FP}
\end{equation}
\begin{equation}\label{recall}
    R=\frac{TP}{TP+FN}
\end{equation}
\begin{equation}\label{f1}
    f_{1}=2*\left(\frac{1}{\frac{1}{P}+\frac{1}{R}}\right)=\frac{2TP}{2TP+FP+FN}
\end{equation}
where TP, FP, and FN denote True Positive, False Positive, and False Negative respectively. The model's Accuracy (Eqn. \ref{accuracy}) indicates the total number of correctly classified instances (both positive and negative classes) from the total number of instances on which the model was applied. The harmonic mean of the precision and recall is denoted by the f1-score which indicates the overall efficiency of the model in predicting only the positive class. A high value of precision means the model predicts fewer FPs while recall indicates the rate at which the model predicts the positive class of objects. In this work, WR sources belong to the positive class whereas the negative class represents the non-WR type objects.
\subsection{Feature significance}\label{subsec:feature_selection}
The important features used in our work are decided from the values of the individual performance metrics (from eqn.\ref{accuracy}-\ref{f1}). Firstly, we develop classifier models for each of the features and identify which feature achieves the highest f1-score (eqn.\,\ref{f1}). To find the optimal set of features that are significant for the f1-score of the classifier, we use the \textit{Recursive Feature Elimination with Cross-Validation} (RFECV$\footnote{\url{https://scikit-learn.org/stable/modules/generated/sklearn.feature_selection.RFECV.html}}$) method. To prevent data leakage, we define a pipeline that fits this \textit{RFE} function on the oversampled 5-randomized subsets (using the \textit{StratifiedKfold}$\footnote{\url{https://scikit-learn.org/stable/modules/generated/sklearn.model_selection.StratifiedKFold.html}}$ technique) of the training dataset and calculate the average f1-score. In brief, \textit{StratifiedKfold} module (from \textit{scikit-learn}) randomly splits the dataset such that the ratio between positive and negative classes of objects remains the same in each of the subsets of the dataset. The classifier model predicts the best set of features from the cross-validation scores generated across all the k-fold subsets. In brief, cross-validation repeatedly splits the dataset into a specified number of subsets (in this case k-fold), trains the model on any combination of k-1 subsets, and tests the model on any $k^{th}$ subset. This approach helps reduce overfitting by confirming that the model performs consistently well across different data segments, rather than relying on a single train-test split. \\
 %Previously, distance based algorithms such as K-Nearest neighbour (KNN) and Support Vector machine (SVM) have been widely used for classification tasks. However, in this paper we make use of the tree based models such as XGB and RF. 

\subsection{Hyper-parameter tuning}\label{subsec:hyper_parameter}
Our classifier (based on the significant features) is trained and cross-validated on randomly oversampled subsets (using the \textit{StratifiedKfold} approach) of the TrD. Using \textit{optuna} \citep{akiba2019optuna} we perform a Bayesian-optimization and optimize the f1-score. The best set of hyper-parameters are derived from a grid of model parameters that achieve the highest f1-score. 
Finally, another classifier model is defined using the determined set of model hyper-parameters and based on the significant features. This final classifier is fitted on the complete oversampled TrD. Thereafter, this model is applied to the TsD, and the labels are predicted.
%%%%%%%%%%%%%%%%%%%%%%%%%%%%%%%%%%%%%%%%%%%%%%%%%%%%%%%%%%%%%%%%%%%%%%%%%%%%%%%%%%%%%%%%%
\section{Results}\label{sec:results}
\subsection{WR versus non-WR classification}\label{subsec:wr_classifier}
\begin{figure*}
    \gridline{\fig{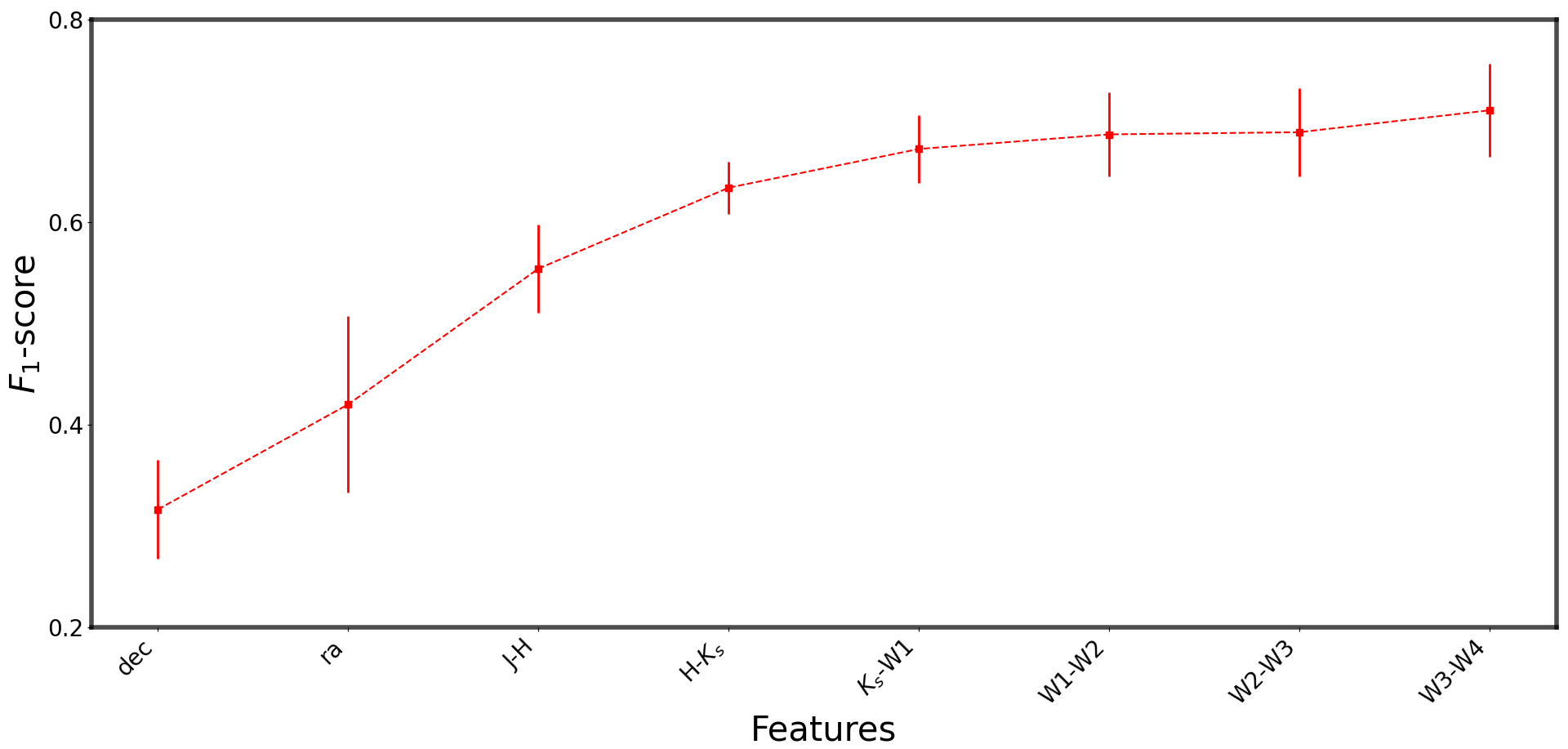}{0.8\textwidth}{(a) RF}
         }
    \gridline{\fig{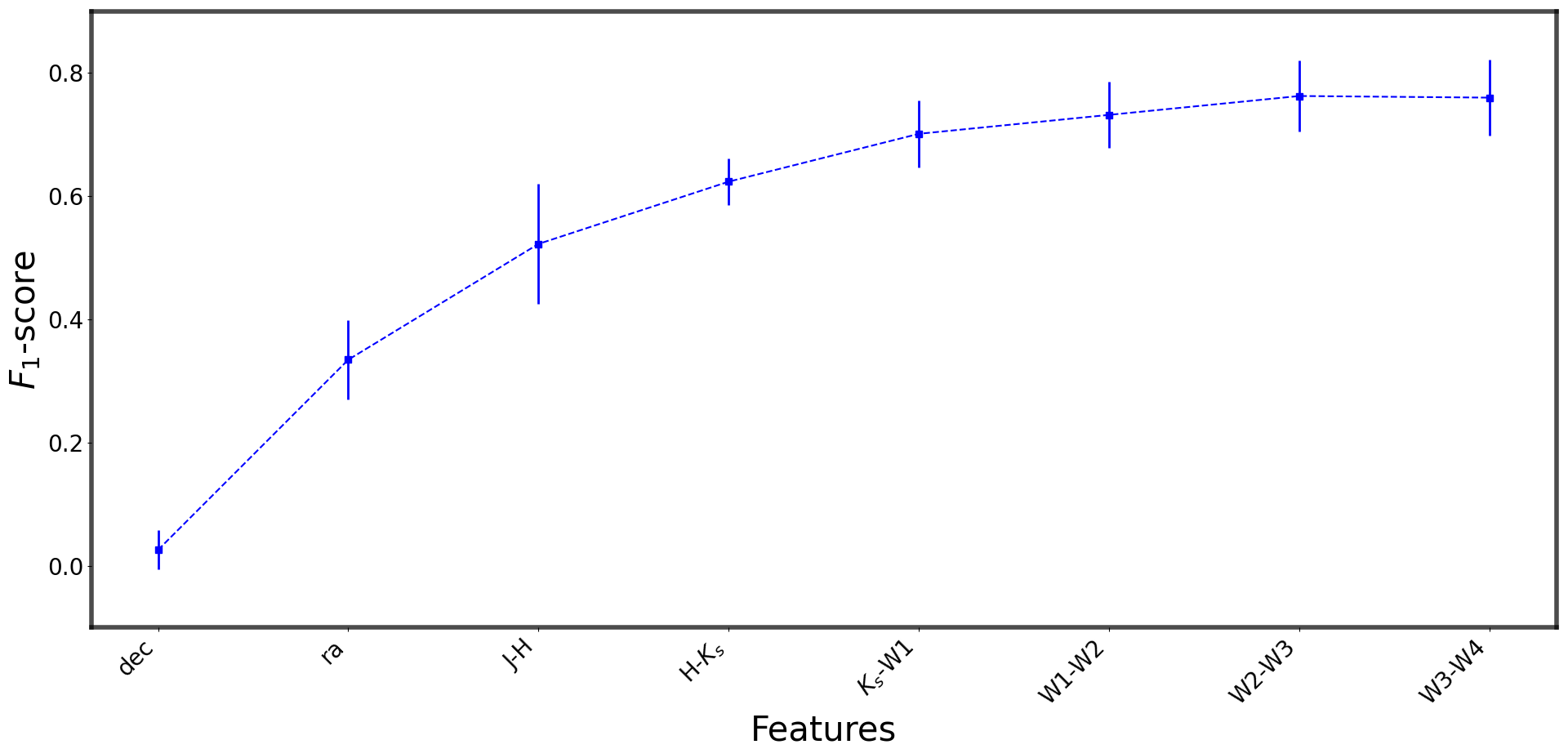}{0.8\textwidth}{(b) XGB}
         }
    \caption{f1-scores of the models with successive combinations of features (from left to right) show the increase in the model's prediction efficiency and eventually reach their highest values with these 8 significant features.}
   \label{fig:feature_importance}
\end{figure*}

Using the RFECV method (as discussed in Sec.\,\ref{subsec: xgboost_model}) the optimal number of features of the classifier model is determined. In Fig.\,\ref{fig:feature_importance}, we show the RFECV curves with our set of features.

\begin{figure*}
    \centering
    \includegraphics[width=1.0\textwidth]{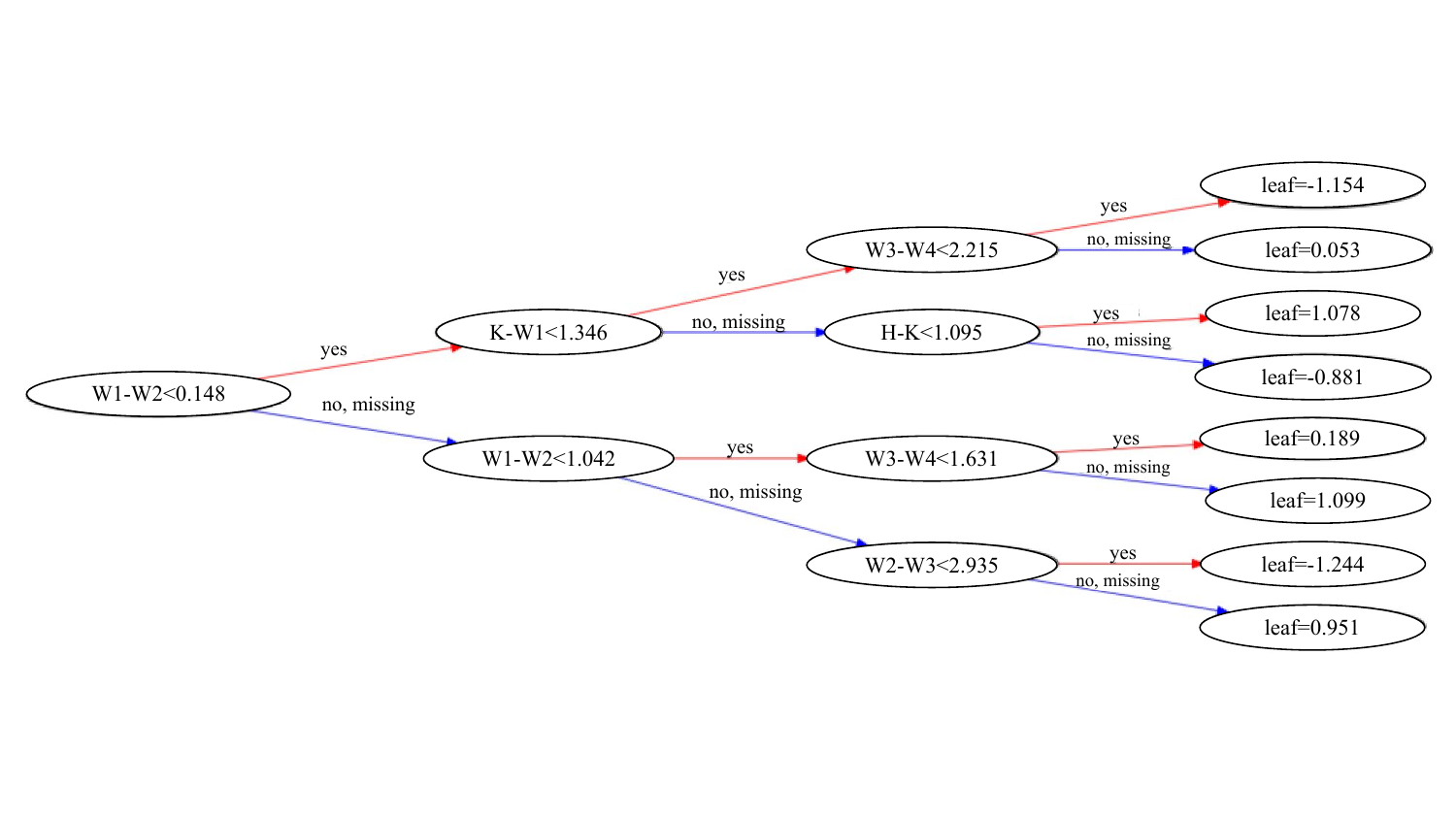}
    \caption{Plot showing the first Decision tree (learner) of the employed XGBoost model. The \textit{internal nodes} store the features which are \textit{branched} based on a criterion and the final predictions are stored in the \textit{leaf nodes}.}
    \label{fig:XGB_model}
\end{figure*}

It is seen that with the sequential addition of more features, the mean of the f1-score (across the Stratified 5-fold randomized subsets of Dataset-1) increases. The mean f1-score saturates beyond W3-W4, which means the model reaches its maximum predictive ability using only these 8 features. Therefore, models based on these features and arranged in the same order (as shown in Fig.\,\ref{fig:feature_importance}) achieve the highest f1-scores. Hence, we use these features to define another XGB and RF classifier which are trained over 5-randomly oversampled subsets of TrD-1 to find the best set of hyper-parameters (see Table\,\ref{tab:model_params}) using Bayesian Optimization technique. Figure \ref{fig:XGB_model} illustrates the first decision tree with model features and classifying conditions of the final XGB model. During the training process, the decision tree automatically chooses and updates the order of the features based on their significance in the classification of objects. Also, the model works on the principle that if a feature value of an object satisfies the selection criteria ('yes'), it is passed on to the next internal node while if it does not ('no', in our case), then it is passed on to some other relevant internal node.

\begin{figure*}
    \centering \includegraphics[width=0.6\textwidth]{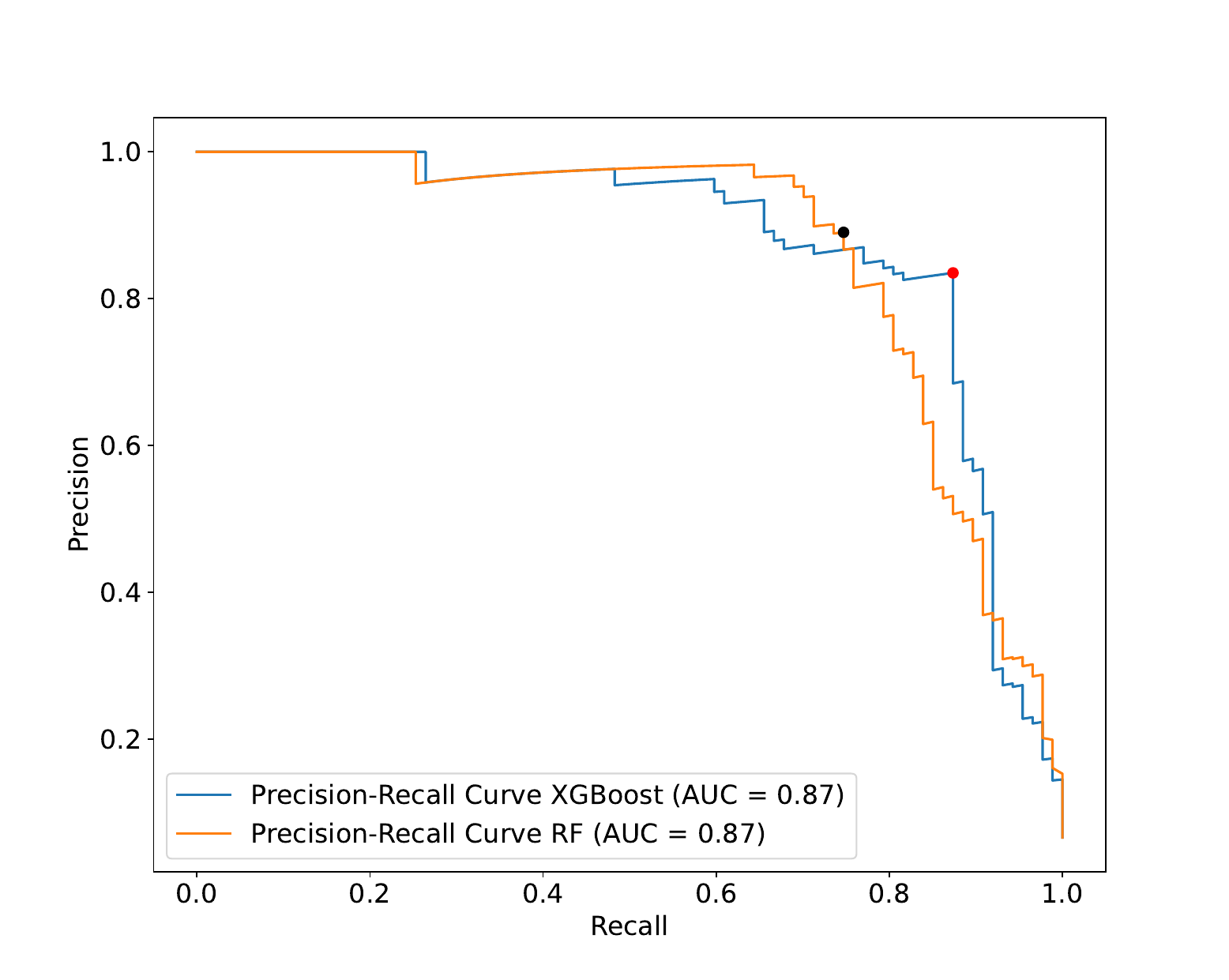}
    \caption{Precision-Recall curve for the optimized models for the classification of the objects. Both models have identical AUCs, but XGB misclassifies fewer WR stars (i.e. higher recall) compared to RF, as indicated by slower growth of recall of RF against XGB while compromising their precisions. The black and red points mark the maximum f1 score achieved by the RF and XGB models respectively showing a higher f1-score for XGB than RF.}
    \label{fig:PRC_curve}
\end{figure*}
\begin{figure*}
\gridline{\fig{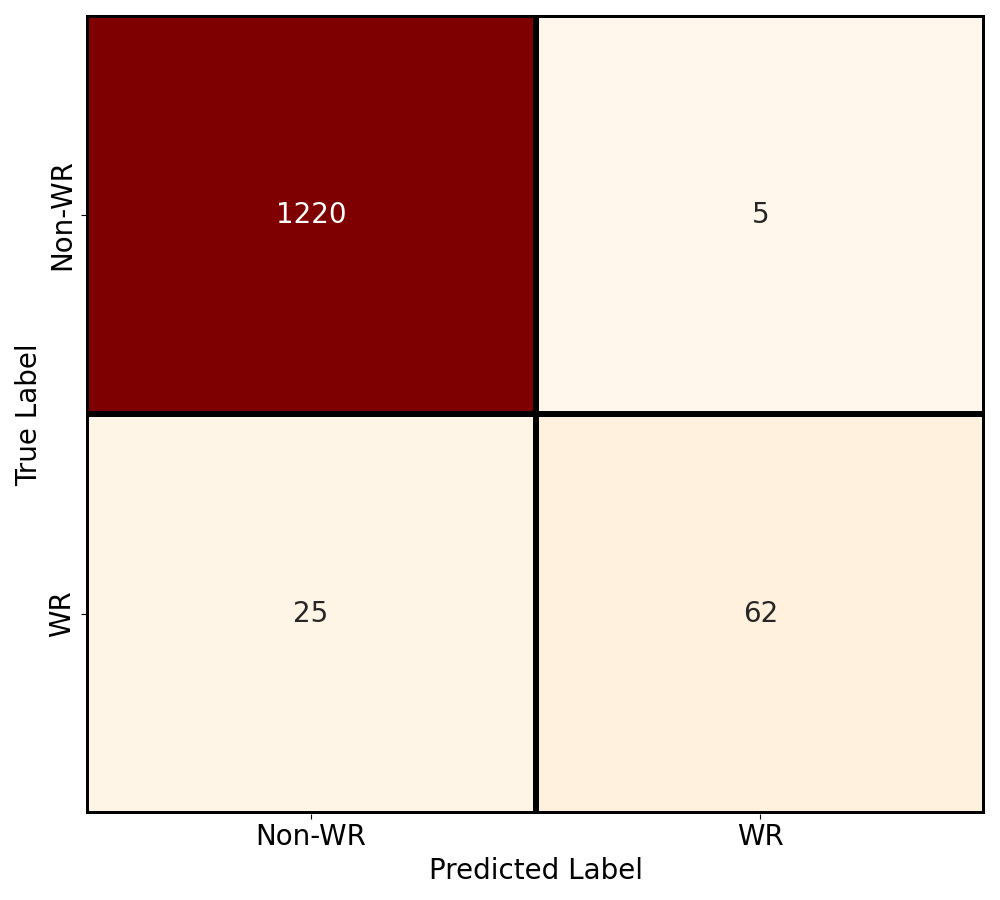}{0.45\textwidth}{(a)}
         \fig{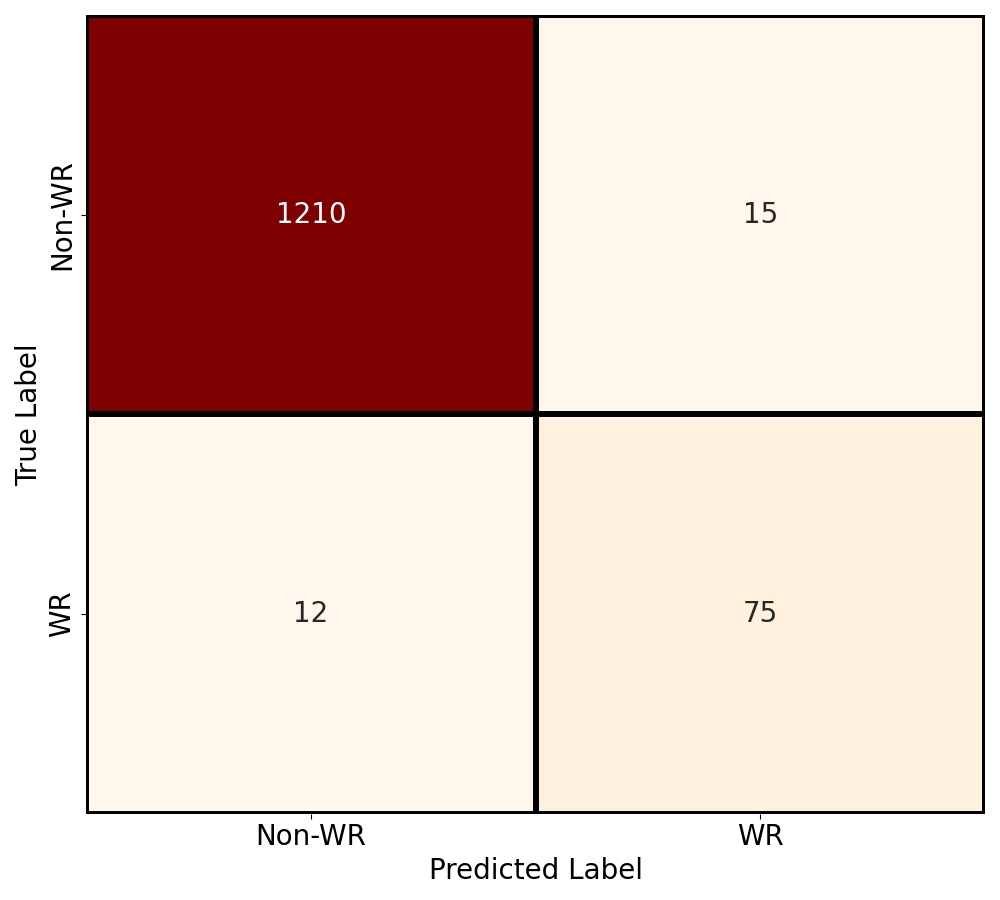}{0.45\textwidth}{(b)}
         }
\caption{Confusion matrices derived from the application of the (a) RF classifier and (b) XGB classifier on TsD-1. The true labels (on the y-axis) indicate the actual class while the model predictions are labeled on the x-axis of the diagrams (see Sec.\,\ref{subsec:wr_classifier} for further information).\label{fig:confusion_matrix_otype}}
\end{figure*}

\begin{deluxetable}{ccc}
\tabletypesize{\scriptsize}
\tablewidth{0pt} 
%\tablenum{4}
\tablecaption{Final model parameters for the Object classifiers.\label{tab:model_params}}
\tablehead{
\colhead{Model} & \colhead{Parameters} & \colhead{Values}
} 
\colnumbers
\startdata 
XGB & $n\_estimator$ & 147 \\
{ } & $max\_depth$ & 3 \\
{ } & $learning\_rate$ & 0.66 \\
{ } & $min\_child\_weight$ & 2 \\ 
{ } & $subsample$ &  0.587 \\ \hline
RF & $n\_estimator$ &  130 \\
{ }  & $max\_depth$ &  None\\
{ } & $max\_features$ & sqrt \\
{ } & $min\_samples\_split$ & 5 \\
{ } & $min\_samples\_leaf$ & 2 \\
\enddata
\end{deluxetable}
\begin{deluxetable}{ccc}
\tabletypesize{\scriptsize}
\tablewidth{0pt} 
%\tablenum{4}
\tablecaption{Performance statistics of Object classifier models applied on the TsD-1.\label{tab:obj_metrics}}
\tablehead{
\colhead{Metric} & \colhead{XGB model} & \colhead{RF model}}
\colnumbers
\startdata 
Accuracy & 98  & 98 \\
Precision & 83 & 92\\
Recall & 86 & 71 \\ 
f1 & 84 & 80 \\
\enddata
\end{deluxetable}
We derive the Precision-Recall (PR) curve (see Fig.\,\ref{fig:PRC_curve}) to compare the performance of the classifiers. The high precision of the RF classifier shows that it detects fewer FPs than the XGB classifier. The Area Under the Curve (AUC) in the case of the XGB classifier is found to be the same as that of the RF classifier. The maximum f1-scores (marked in the plot) achieved by the models indicate that the XGB performs better than the RF model. This is also seen from Table\,\ref{tab:obj_metrics}.
We also show the performance of the classifiers using these confusion matrices (see Fig.\,\ref{fig:confusion_matrix_otype}); which provide a detailed breakdown of the model's predictions by comparing the actual and predicted classes. The diagonal elements represent the true predicted instances and the remaining elements indicate the wrong classifications. We note that the XGB classifier (Fig.\,\ref{fig:confusion_matrix_otype}(b)) outperforms the RF-classifier (Fig.\,\ref{fig:confusion_matrix_otype}(a)) in the accurate predictions of the minority class. We further compare their performances in Sec.\,\ref{sec:discussion}.

\begin{figure*}
\gridline{\fig{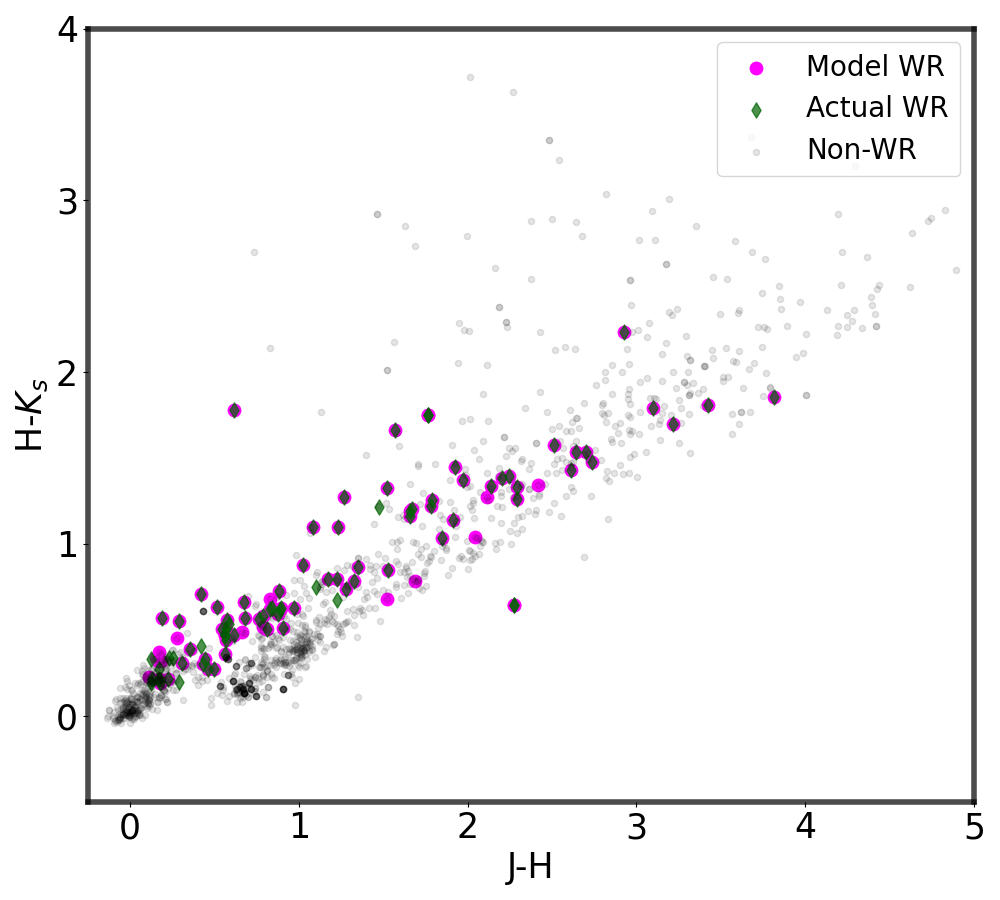}{0.5\textwidth}{(a)}
          \fig{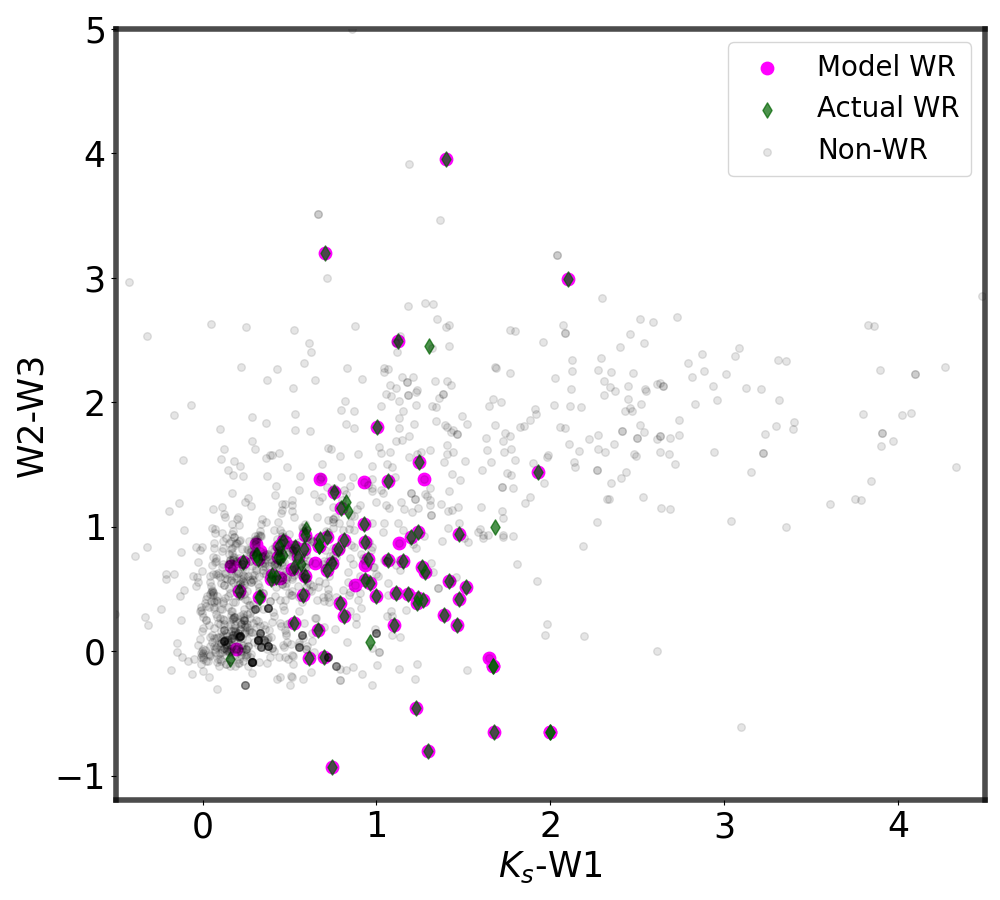}{0.5\textwidth}{(b)}
         }
\gridline{\fig{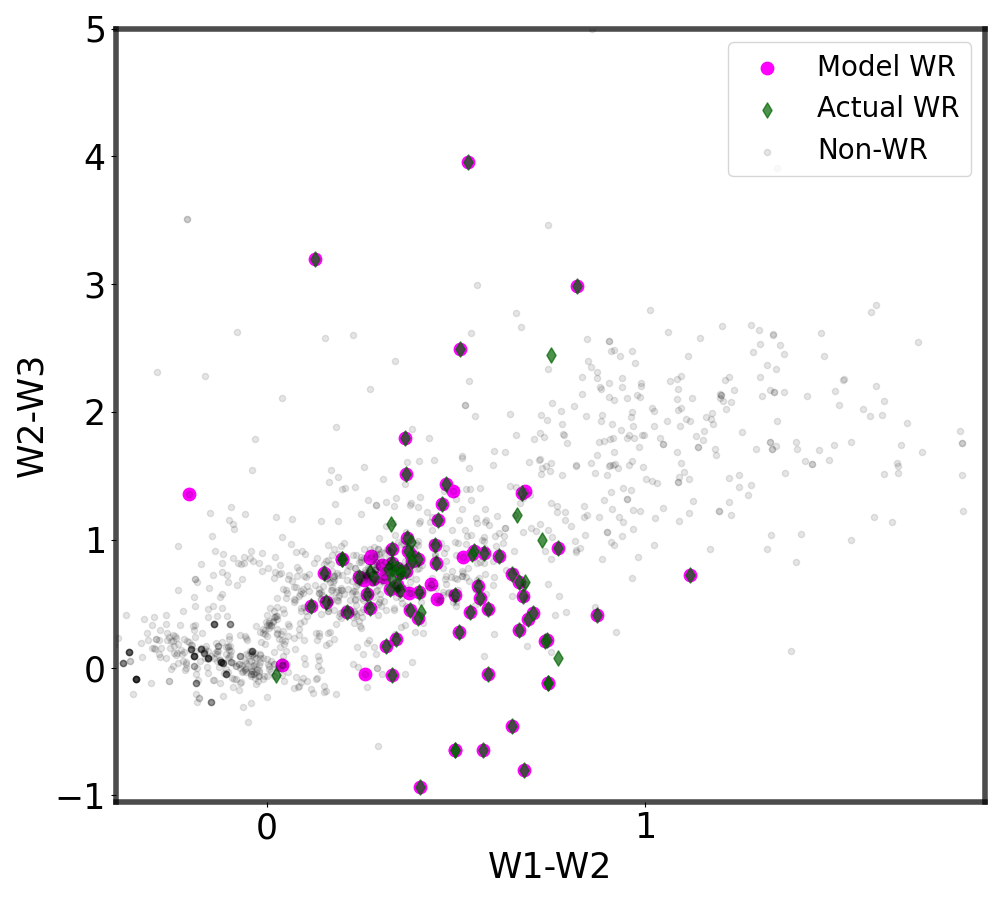}{0.5\textwidth}{(c)}
          \fig{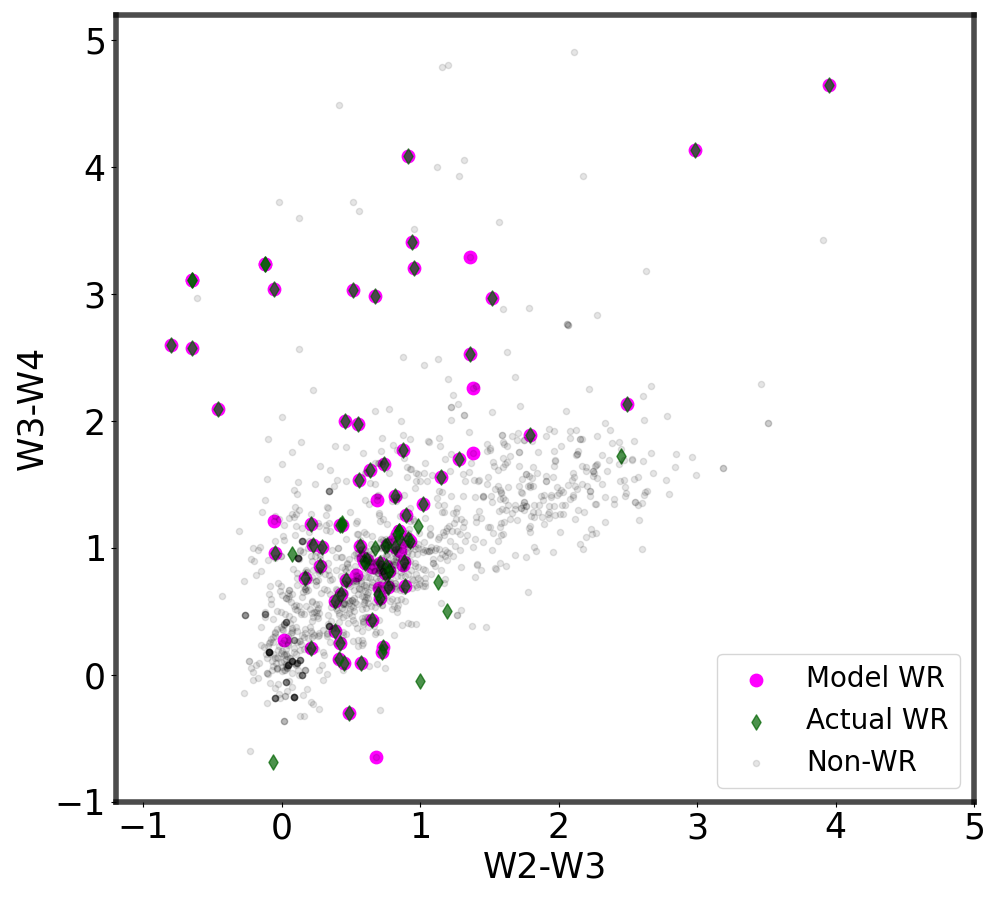}{0.5\textwidth}{(d)}
          }
\caption{Predictions by the best XGB classifier on the TsD-1. \textit{Grey} circles represent non-WR candidates (that are mainly the Be-type and AGB sources), while the \textit{green} diamonds and \textit{magenta} circles represent the actual and model WR candidates respectively. The overlapping symbols (i.e. the green-colored diamonds and magenta-colored circles) represent the TPs.} \label{fig:xgb_prediction}
\end{figure*}

\begin{figure*}
\gridline{\fig{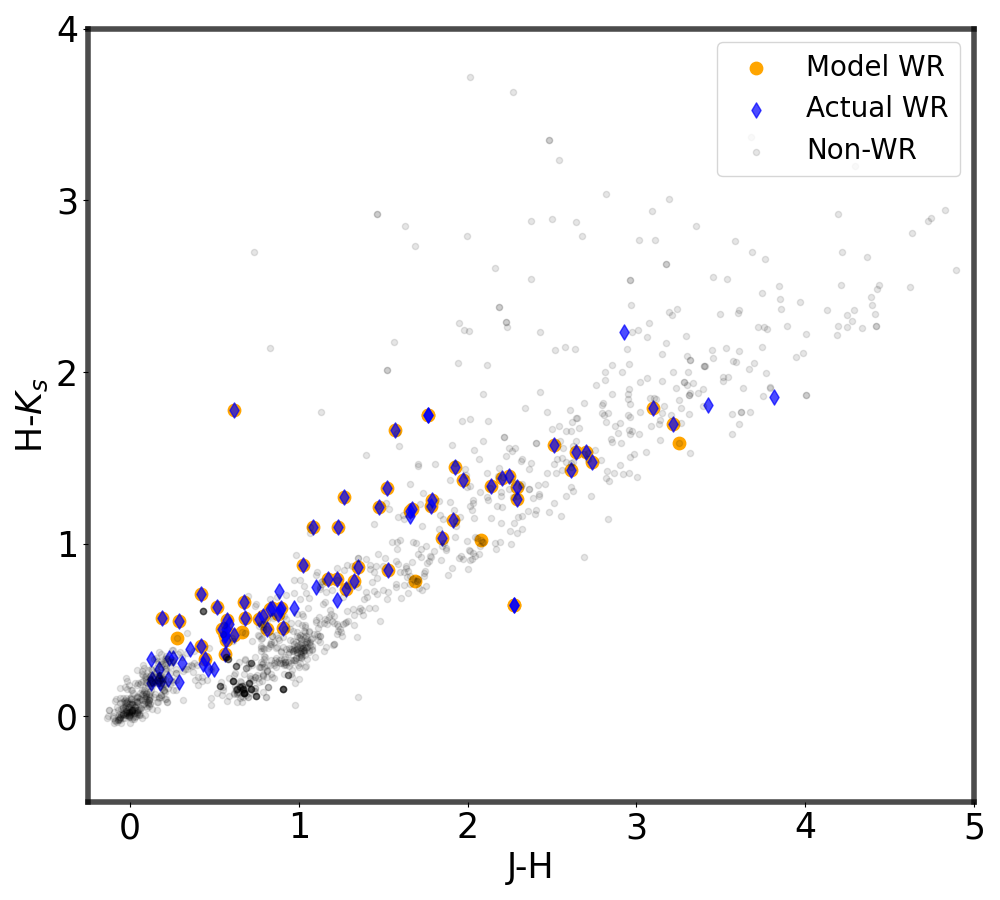}{0.5\textwidth}{(a)}
          \fig{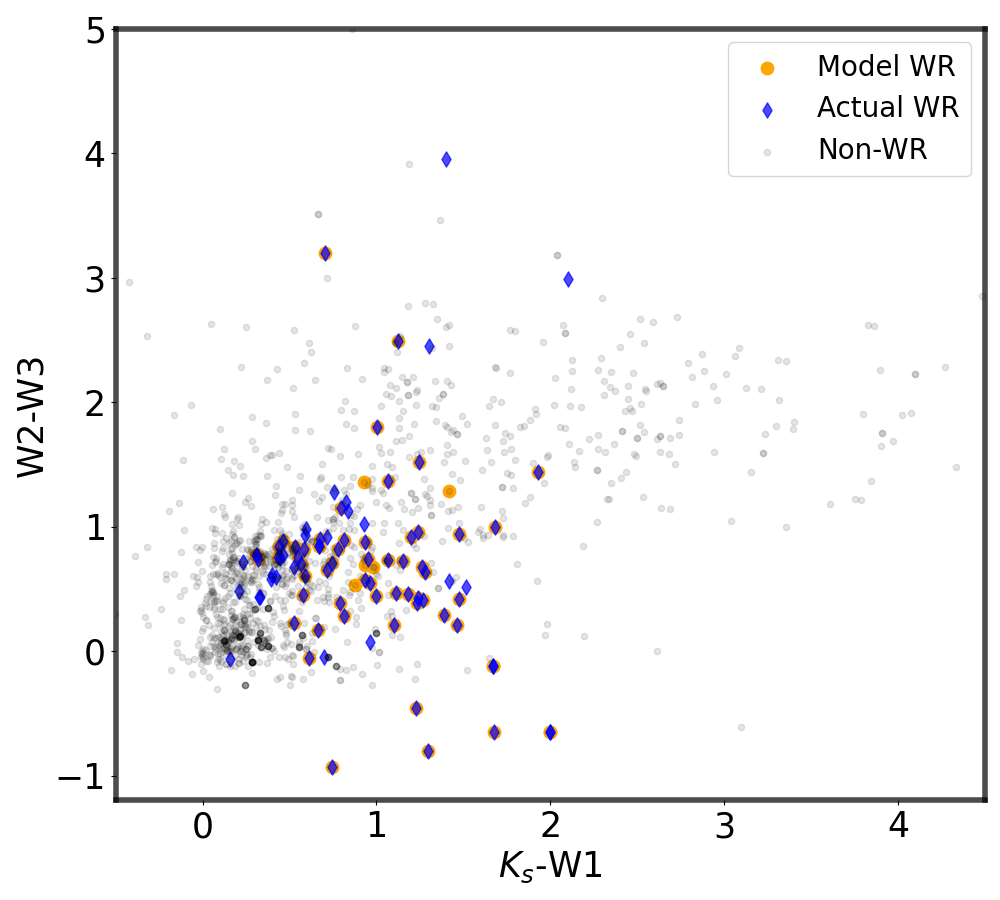}{0.5\textwidth}{(b)}
         }
\gridline{\fig{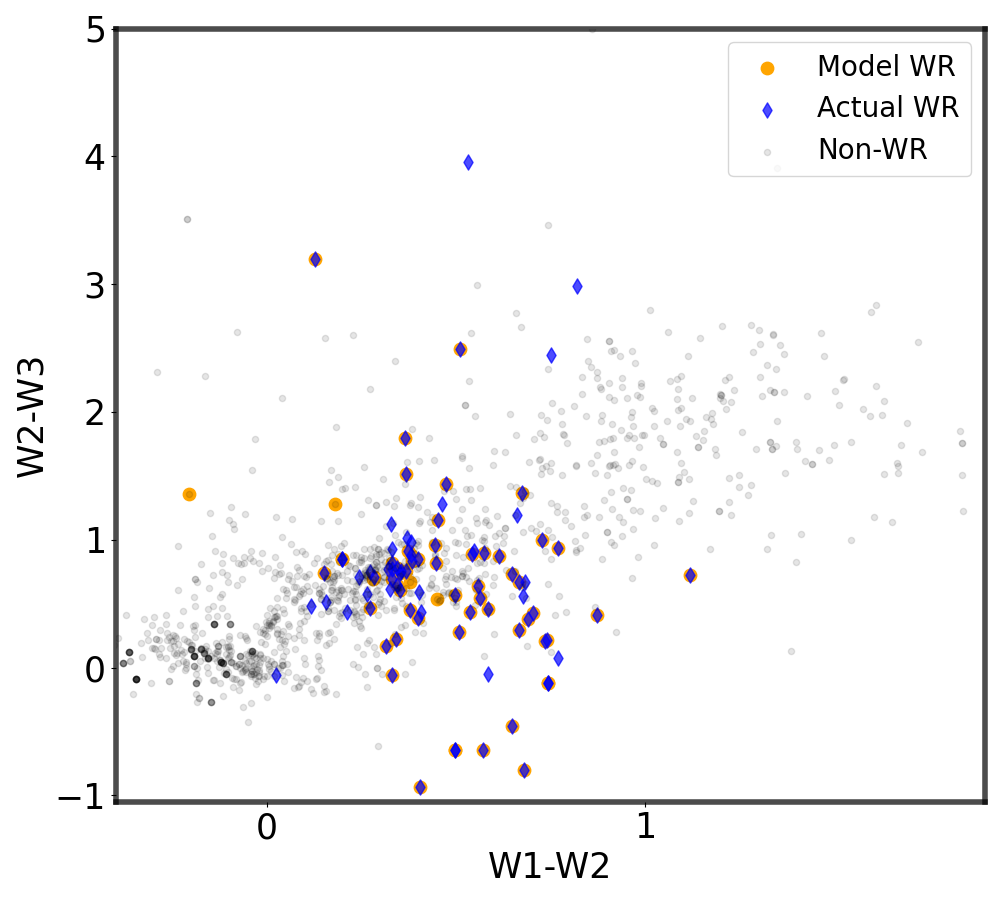}{0.5\textwidth}{(c)}
          \fig{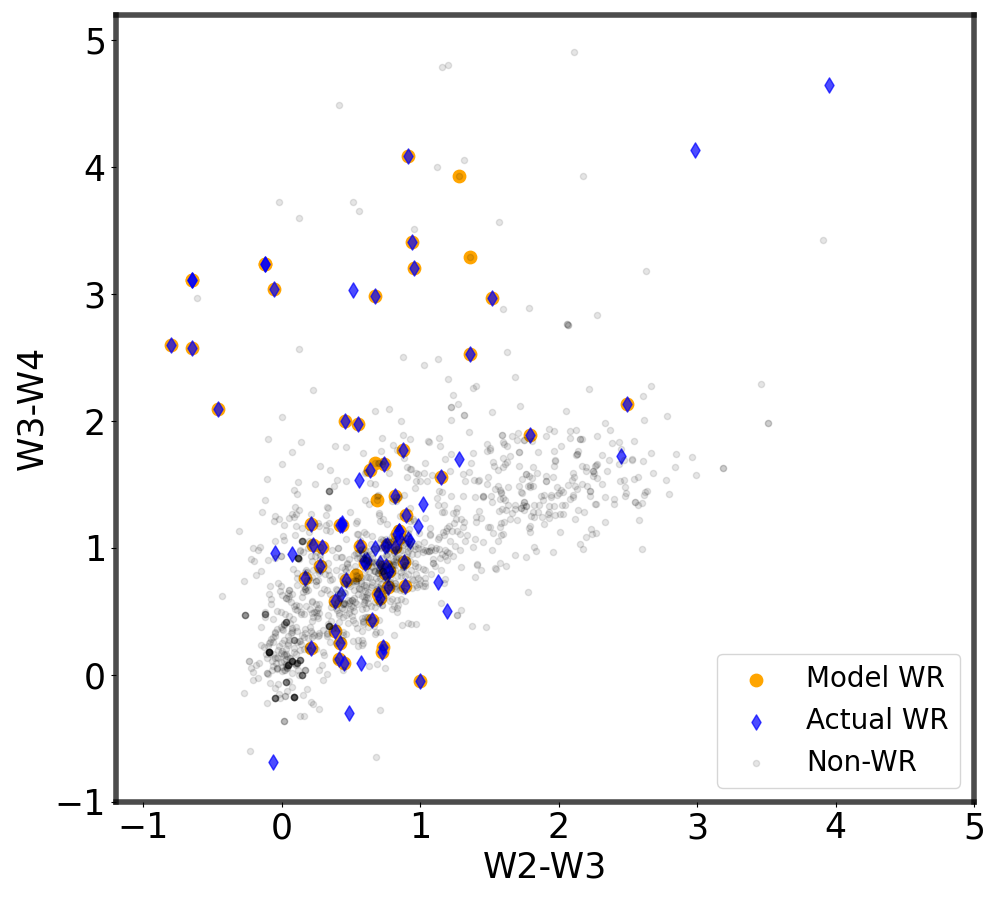}{0.5\textwidth}{(d)}
          }
\caption{Predictions by the best RF classifier on the TsD-1. \textit{Grey} circles represent non-WR candidates (same as in Fig.\,\ref{fig:xgb_prediction}), while the \textit{blue} diamonds and \textit{orange} circles represent the actual and model-predicted WR candidates respectively. The TPs are identified by overlapping blue diamonds and orange circles.} \label{fig:rf_prediction}
\end{figure*}

In Figures \ref{fig:xgb_prediction} and \ref{fig:rf_prediction}, we show the color-color predictions of TPs by the final classifiers when applied to the test dataset (TsD-1). In our datasets, the AGB and Be-type stars are the dominant classes which are known for IR colors similar to that of WR stars. The NIR color space of classical Be-type stars (with free-free emission from the ionized circumstellar disks) overlaps with that of the WR stars and is detected by the models (see Fig.\,\ref{fig:xgb_prediction}(a) and \ref{fig:rf_prediction}(a)) as FPs that mostly lie in the clustered region of the color spaces \citep{2011IAUS..272..404L}. The Spectral Energy Distribution (SED) of WR stars with color-excess due to strong free-free emission and/or circumstellar dust in general peaks around the W2-band \citep{2023ApJ...951...89L}, coinciding with those of AGB stars with dust emission \citep{suh2020infrared}. This overlap results in the poor detection statistics of WR stars in the W2-W3 color space (see Fig.\,\ref{fig:xgb_prediction}(b) and \ref{fig:rf_prediction}(b)). However, at longer MIR bands, the dust emission from the AGBs is significantly less pronounced than that from WR stars, making WR stars more easily identifiable in the W3-W4 color space (see Fig.\,\ref{fig:xgb_prediction}(d) and \ref{fig:rf_prediction}(d)). This phenomenon can be understood from the context of the circumstellar envelopes for WR stars and those of AGB. WR stars experience higher mass loss rates and have larger radii than AGB stars, leading to more extended circumstellar envelopes. These envelopes emit strongly, causing noticeable excesses in various color bands. AGB stars also show similar excesses due to their dusty circumstellar envelopes. However, this effect is absent in the W3-W4 band because of their less extended outer shells.

\subsection{WR sub-type classification}\label{subsec:wr_subtype}
%In the MWG, the WC subtypes a
From the previous section, we can conclude that the XGB model is a highly efficient classifier model. Therefore, we develop a multi-class XGB classifier model to identify the WR sub-types (WC and WN-types) from the non-WR objects. We follow the same feature selection approach using the RFCEV method (as discussed in \ref{subsec:feature_selection}) to find the important features required by the model to achieve the best f1-score. As this has more than two classes, we use the macro f1-score which is the average of f1-scores for different classes. Thereafter, we determine the optimal parameters (using the Bayesian optimization technique) upon training and cross-validating the model on 5 randomized subsets of Dataset-2 (with known spectral types) and optimize the macro f1-score. The final classifier (Table \ref{tab:sub_model_params}) is fitted on the oversampled training dataset (Trd-2) and tested on Tsd-2 to determine the model predictions.

\begin{deluxetable}{ccc}
\tabletypesize{\scriptsize}
\tablewidth{0pt} 
%\tablenum{4}
\tablecaption{Final model parameters for the WR subtype classifier.\label{tab:sub_model_params}}
\tablehead{
\colhead{Model} & \colhead{Parameters} & \colhead{Values}
} 
\colnumbers
\startdata 
XGB & $n\_estimator$ & 131 \\
{ } & $max\_depth$ & 3 \\
{ } & $learning\_rate$ & 0.305 \\
{ } & $min\_child\_weight$ & 3 \\ 
{ } & $subsample$ & 0.529 \\ \hline
\enddata
\end{deluxetable}
\begin{figure*}
    \centering \includegraphics[width=0.45\textwidth]{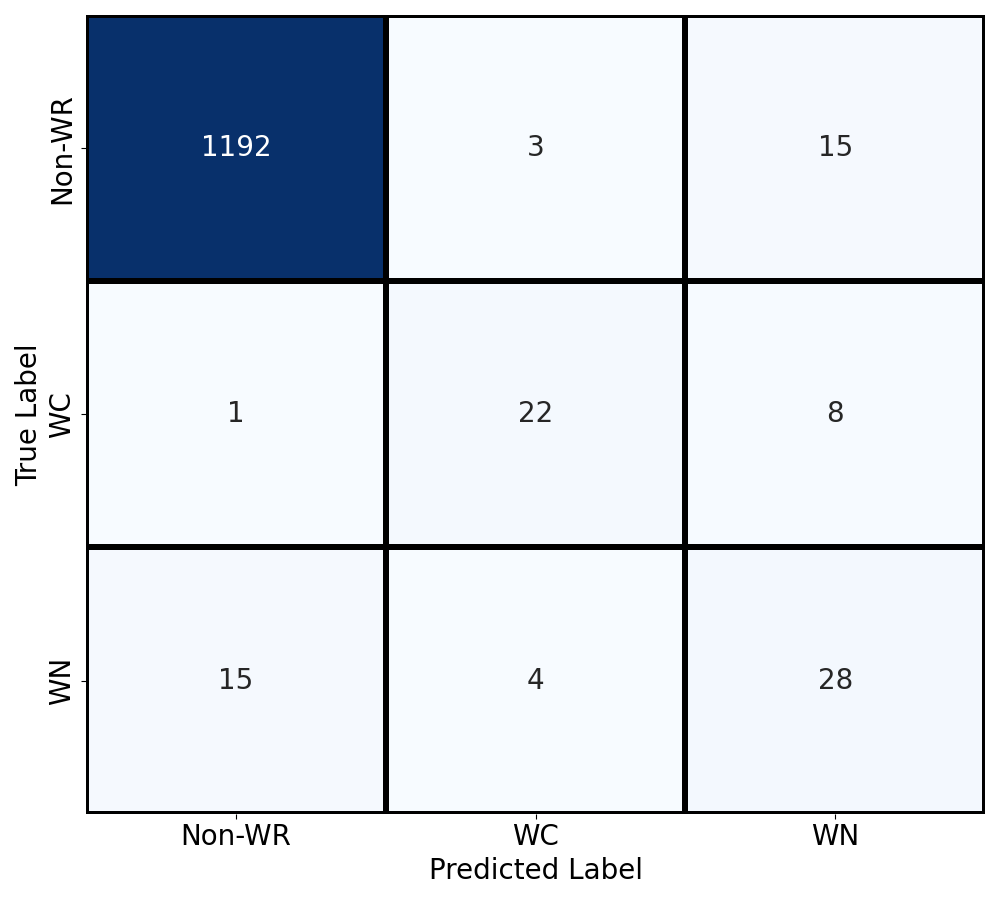}
    \caption{Confusion matrix derived from the WR-subtype classifier predictions (see Sec.\,\ref{subsec:wr_subtype}).}
    \label{fig:xgb_subtype_cm}
\end{figure*}
\begin{deluxetable}{ccc}
\tabletypesize{\scriptsize}
\tablewidth{0pt} 
%\tablenum{4}
\tablecaption{Performance statistics of WR subtype classifier model applied on the TsD-2.\label{tab:subtype_metrics}}
\tablehead{
\colhead{Metric} & \colhead{WC-class} & \colhead{WN-class}}
\colnumbers
\startdata 
Precision & 76 & 55 \\
Recall & 71 & 60 \\ 
f1 & 73 & 57 \\
\enddata
\end{deluxetable}
The prediction of the model on the TsD-2 is shown in the form of a confusion matrix (see Fig.\,\ref{fig:xgb_subtype_cm}). From Table\,\ref{tab:subtype_metrics}, we find that the WC and WN subtypes are detected with a recall of 71\% and 60\% respectively. A lower number of misclassifications between WC and WN with an overall high detection accuracy ($\sim 97\%$) is a strong indicator of the model's ability to accurately distinguish between WR sub-types. We find more misclassification of non-WR sources (i.e. FPs) as WN-type stars compared to WC-type stars. Such non-WR sources are mostly the classical Be-type sources \citep{2011IAUS..272..404L} which contaminate the color space of WN-type stars. The WC-type stars show color excess due to strong carbon emission lines and sometimes due to circumstellar dust in the $K_{s}$-band \citep{2011AJ....142...40M} making them distinguishable from the dominant classes (i.e., Be and AGB) of non-WR sources. 

\begin{figure*}
     \centering \includegraphics[width=0.7\textwidth]{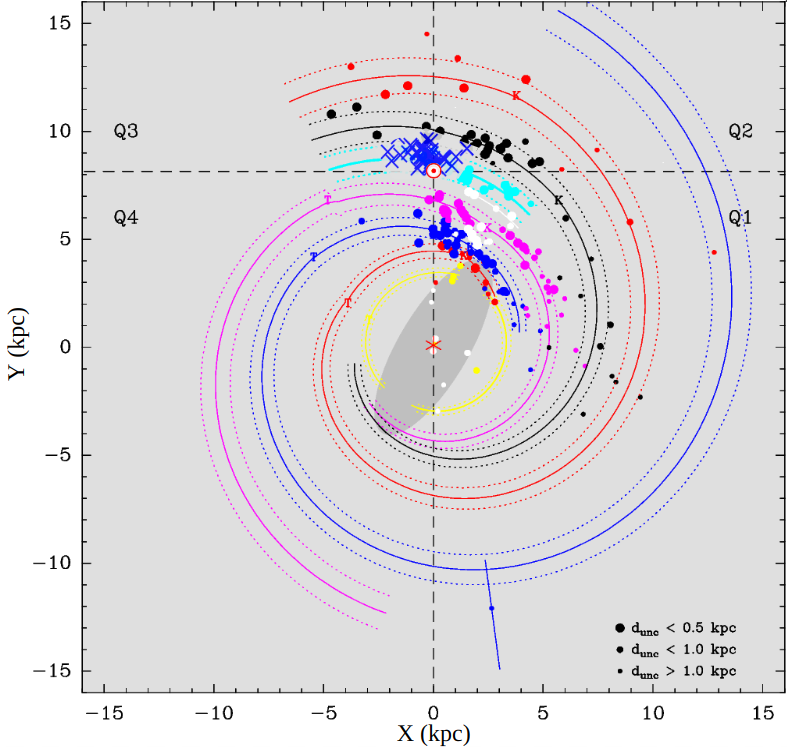}
    \caption{Galactic distribution of possible WR star candidates (marked as \textit{blue} cross) predicted by the XGB-object type classifier (see Sec.\,\ref{subsec:new_wr}). The dataset is over-plotted on the Galactic plane showing massive star formation zones across different arms of the MW (for detail, see \citet{reid2019trigonometric}).}
    \label{fig:new_wr}
\end{figure*}
\begin{deluxetable*}{lccccccccccc}
\tablecaption{Catalogue of predicted WR stars (see Sec. \ref{subsec:new_wr}) \label{tab:new_wr_cat}}
\tablehead{
\colhead{Source} & \colhead{Sub-type} & \colhead{RA} & \colhead{Dec} & \colhead{J} & \colhead{H} & \colhead{$K_{s}$} & \colhead{W1} & \colhead{W2} & \colhead{W3} & \colhead{W4} & \colhead{Parallax}\\
\colhead{(2MASS)} &  & \colhead{(hh:mm:ss)} & \colhead{(dd:mm:ss)} & \colhead{(mag)} & \colhead{(mag)} & \colhead{(mag)} & \colhead{(mag)} & \colhead{(mag)} & \colhead{(mag)} & \colhead{(mag)} & \colhead{(mas)}
}
\startdata
J00463205+6357057 & WC  & 00:46:32 & +63:57:05 & 9.711  & 9.442  & 8.411  & 8.937  & 9.059  & 9.036 & 8.365 & 4.167$\pm$0.022\\
J08493645-4132202 & WN  & 08:49:36 & -41:32:20 & 14.465 & 13.540 & 13.054 & 12.390 & 12.092 & 9.482 & 6.409 & 1.206$\pm$0.076\\
J10443325-5925084 & WN  & 10:44:33 & -59:25:08 & 12.171 & 11.777 & 11.247 & 10.361 & 10.055 & 6.878 & 2.462 & 0.435$\pm$0.018\\
J15314250-5158294 & WC  & 15:31:42 & -51:58:29 & 6.768  & 7.778  & 5.617  & 4.641  & 4.518  & 4.631 & 4.537 & 1.133$\pm$0.059\\
J16162732-5433377 & WC  & 16:16:27 & -54:33:37 & 12.239 & 11.875 & 10.635 & 9.443  & 9.624  & 9.387 & 8.386 & 1.459$\pm$0.020\\
J18201030-1613461 & WC  & 18:20:10 & -16:13:46 & 12.217 & 11.089 & 10.155 & 8.895  & 8.197  & 8.323 & 0.754 & -\\
J19155721+0213379 & WN  & 19:15:57 & +02:13:37 & 7.979  & 7.452  & 7.220  & 7.114  & 7.207  & 7.157 & 7.116 & 1.893$\pm$0.020\\
J19475176+2414284 & WN  & 19:47:51 & +24:14:28 & 12.753 & 11.767 & 11.290 & 10.651 & 10.249 & 7.090 & 5.076 & 0.592$\pm$0.041\\
J19492779+2743394 & WN  & 19:49:27 & +27:43:39 & 9.814  & 9.207  & 8.821  & 8.076  & 7.961  & 7.111 & 6.251 & 0.947$\pm$0.369\\
J20375570+5106184 & WN & 20:37:55 & 51:06:18 & 8.931 & 8.435 & 8.361 & 7.231 & 6.772 & 7.259 & 7.204 & 14.448$\pm$0.011\\
\enddata
\tablecomments{Complete catalogue of predicted WR stars is available in a machine readable format. Here we have shown only those WR candidates that have been identified by both the object and subtype models.}
\end{deluxetable*}
\subsection{New WR star identification}\label{subsec:new_wr}
Finally, we apply our models on a large and different dataset from SIMBAD comprising stars with unknown object types. For this, we retrieved 18750 stellar objects (restricted to Galactic sources using the same positional criteria as mentioned in Sec.\,\ref{subsec:data}) from SIMBAD with no information on their object types. The data is sorted in the same manner (i.e. objects with available photometric magnitudes and present in the same color space) as mentioned in Sec.\,\ref{subsec:data} and apply the XGB object type classifier model on the reduced dataset (comprising 6457 objects). The model classifies 58 objects as WR stars (see Table\,\ref{tab:new_wr_cat}). Also, we apply the subtype classifier to the same dataset to identify the subtypes. Among the predicted list of objects, the model identifies the subtypes of 10 of the 58 WR star candidates detected by the object classifier (see Table\,\ref{tab:new_wr_cat}). 
We also obtain the parallaxes for the WR star candidates by cross-matching them with Gaia DR3 \citep{2016A&A...595A...1G, 2023A&A...674A...1G} and find that except two (with missing parallax information), all of them are located within 2\,kpc from the sun. This can be seen in Fig.\,\ref{fig:new_wr}, where we overplot the WR candidates on a 2D X-Y plot from \citet{reid2019trigonometric} showing all the Galactic arms. We find that most of our detected sources lie on the Local arm with a few located on the Perseus arm. This is expected as the Local arm is known to be a high-mass star-forming region and WR stars are very likely to be found in these regions \citep{2015MNRAS.447.2322R}. Given the spatial distributions of our WR candidates, we expect a majority of them to be actually WR stars. However, there is a chance of FPs in the detected list of WR candidates as the dataset may contain objects which differ from the training data (i.e. weaker sources or weak/high extinction). Therefore, a spectroscopic follow-up of these objects is suggested to confirm our results. The above dataset may also contain various other types of bright astronomical objects that may not have been considered (such as OB-type stars, Compact H\,$\mathbf{\textsc{ii}}$ regions, etc.) during our model development which may affect the detection of WR stars. We plan to include them in future work.

\section{Discussion}\label{sec:discussion}
We compare our classification results with those of \citet{2011AJ....142...40M}. From Fig.\,\ref{fig:xgb_prediction} and Fig.\,\ref{fig:rf_prediction}, we note that our models can identify WR stars across the broad IR color space with enhanced accuracy than what was achieved from selecting WR stars from so-called sweet-spot in the color-color diagrams. Further, our models can detect WR stars with much greater accuracy (see Table\,\ref{tab:obj_metrics}) than detected (50\%) by \citet{2011AJ....142...40M} from the color-based manual identification criteria. Additionally, both of our models (XGB and RF) perform much better than the earlier KNN \citep{2018MNRAS.473.2565M} and SVM \citep{2021ApJ...913...32D} models in accurate label predictions of WR stars. This shows that the ensemble-based algorithms are superior to the instance-based algorithms in classifying stellar candidates from a dataset consisting of large, imbalanced, and heterogeneous objects.

We claim that both of our object classifier models (see Sec.\,\ref{subsec:wr_classifier}) can identify WR stars present across different metallicity regions \citep{2015MNRAS.447.2322R} of the MW. Overall, based on the f1-score (see Table\,\ref{tab:obj_metrics}), we note that the XGB does a slightly better WR-identification task than the RF model. The high recall of the XGB classifier occurs due to the smaller number of FNs detected compared to the RF classifier model. 
However, we find a better f1-score for XGB compared to RF. The XGB-classifier can identify the WR sources lying in the clustered regions of the IR color spaces (see Fig.\,\ref{fig:xgb_prediction}) better than the RF model (see Fig.\,\ref{fig:rf_prediction}).

Both the classifiers predict that (from Table\,\ref{tab:Otype_FPs}) the major contaminants for WR sources in the NIR color space are the Be-type objects which was also reported in \citet{Faherty_2014}. While those in the mid-IR are the O-rich AGBs. In the list of FPs, one of the objects is an HMXB system of WR stars. Therefore, it can be rather considered to be a TP.

\begin{deluxetable*}{lcccccccccc}
\tablecaption{FPs detected by the object classifier models. \label{tab:Otype_FPs}}
\tablewidth{0pt}
\tablehead{
\colhead{Source} & \colhead{Dec} & \colhead{RA} & \colhead{Object type} & \colhead{Spectral type} & \colhead{J-H} & \colhead{H-$K_{s}$} & \colhead{$K_{s}$-W1} & \colhead{W1-W2} & \colhead{W2-W3} & \colhead{W3-W4}\\
\colhead{(2MASS)} & \colhead{(deg)} & \colhead{(deg)} & \colhead{} & \colhead{ } & \colhead{ } & \colhead{ } & \colhead{ } & \colhead{} & \colhead{ } & \colhead{ }
}
\startdata
  & & & & XGB model & & & & & & \\ \hline
 J15401711-5803497 & -58.0638 & 235.0713 & Be & O/Be & 0.166 & 0.319 & 0.33 & 0.304 & 0.802 & 1.056 \\
 J16064773-5225033 & -52.4176 & 241.6989 & AGB & O-rich & 2.416 & 1.341 & 1.128 & 0.519 & 0.865 & 1.046 \\
 J16075617-1859492& -18.9970	& 241.9841	& Mira & - & 0.787	& 0.518	& 0.648	& 0.309 & 0.705 & 0.684 \\
 J10302384-6004404 & -60.0779 & 157.5994 & Be & B5 & 0.115 & 0.227 & 0.164 & 0.257 & 0.681 & -0.648 \\
 J17210447-3502401 & -35.0445 & 260.2686 & AGB & O-rich & 2.043 & 1.041 & 1.650 & 0.259 & -0.054 & 1.215 \\
 J18400705-0525346 & -5.4263 & 280.0294 & RSG & M4Ib & 1.520 & 0.681 & 0.677 & 0.493 & 1.382 & 1.751 \\
 J10441961-5916590 & -59.2831 & 161.0817	& Be & O9.7:V:(n)e & 0.181	& 0.321	& 0.444	& 0.377 & 0.586 & 0.924 \\
 J10294719-5636465 & -56.6129	& 157.4466	& Be & B0Ve & 0.173	& 0.372	& 0.474	& 0.276	& 0.875	& 0.868 \\
 J20322578+4057279 & 40.9578 & 308.1074 & HMXB & WR+WR (WN4/5-6/7) & 2.117 & 1.271 & 1.275 & 0.682 & 1.380 & 2.264 \\
 J18390419-4803184 & -48.0551 & 279.7675 & AGB & M7 & 0.807 & 0.506 & 0.516 & 0.435 & 0.655 & 0.846\\
 J17070515-3545323 & -35.7590 & 256.7715 & Be & B1:IIInne & 0.279 & 0.450 & 0.875 & 0.450 & 0.533 & 0.788 \\
 J15132778-6021388 & -60.3608 & 228.3658 & RSG &	M0Iab-Ib & 0.828 & 0.678 & 0.195 & 0.040 & 0.018 & 0.278 \\
 J11440030-6107364 & -61.1268 & 176.0013 & HMXB & B0.5Ve & 0.665 & 0.486 & 0.932 & 0.282 & 0.690 & 1.381 \\
 J10410408-5834130 & -58.5703 & 160.2670 & Be & Oe & 0.154 & 0.334 & 0.305 & 0.274 & 0.857 & 0.979 \\
 J18233443-1439473 & -14.6631 & 275.8935 & AGB & M6III & 1.686 & 0.783 & 0.930 & -0.206 & 1.361 & 3.289\\ \hline
  & & & & RF model & & & & & & \\ \hline
 J17453819-2917201 & -29.2889 & 266.4091 & AGB & - & 3.251 & 1.587 & 1.419 & 0.180 & 1.283 & 3.932 \\
 J18452531-0323011 & -3.3837 & 281.3555 & RSG &	M3Ia & 2.078 & 1.024 & 0.982 & 0.380 & 0.672 & 1.672 \\
 J17070515-3545323 & -35.7590	& 256.7715	& Be & B1:IIInne & 0.279	& 0.450 & 0.875 & 0.450	& 0.533	& 0.788 \\
 J11440030-6107364 & -61.1268	& 176.0013	& HMXB & B0.5Ve & 0.665	& 0.486	& 0.932	& 0.282	& 0.690 & 1.381 \\
 J18233443-1439473 & -14.6631 & 275.8935 & AGB & M6III & 1.686 & 0.783 & 0.930 & -0.206 & 1.361 & 3.289 
\enddata
\end{deluxetable*}

In Sec.\,\ref{subsec:wr_subtype}, we find that the subtype XGB-classifier model classifies both the WC and WN type objects from the non-WR sources with good accuracy rates. This means that the model can distinguish between the WR subtypes without the need for corresponding narrow-band IR colors or spectroscopic follow-ups. It can be seen (from Table\,\ref{tab:Sptype_TPs}) that among the WC-type stars, most of the TPs detected by the classifier are of WCL-type. This shows that the model can identify the characteristic color-space of these objects that are mainly found in metal-rich galaxies \citep{2007ARA&A..45..177C} such as the MW. The contaminants for WC-class are mostly the WN stars \citep{Faherty_2014} that are mostly early-type while the reverse is not observed (see Table\,\ref{tab:Sptype_FPs} in Appendix~\ref{appendix}). Further, it is noted (from Table\,\ref{tab:Sptype_TPs}) that the TPs corresponding to the WN-type stars span across WNE to WNL-type. However, mostly the WNL-type objects dominate the list as they exhibit color-excess in the W3-W4 space due to free-free emission. The model can also detect the WNh-type objects that are classified as non-classical WR stars. However, we note that the WNL and WCL show similar color patterns leading to the observed misclassifications by the model (see Fig.\,\ref{fig:xgb_subtype_cm}). This is further confirmed by the detected FPs corresponding to the WN-type objects (see Table\,\ref{tab:Sptype_FPs} in Appendix~\ref{appendix}). This happens because of the excess emission due to free-free emission and sometimes due to circumstellar dust formed in earlier stages of evolution \citep{2003Ap&SS.285..677C}.

\begin{deluxetable*}{lccccccccc}
\tablecaption{TPs detected by the Sub-type classifier model.\label{tab:Sptype_TPs}}
\tablewidth{0pt}
\tablehead{
\colhead{Source} & \colhead{Dec} & \colhead{RA} & \colhead{Sub-type} & \colhead{J-H} & \colhead{H-$K_{s}$} & \colhead{$K_{s}$-W1} & \colhead{W1-W2} & \colhead{W2-W3} & \colhead{W3-W4}\\
\colhead{(2MASS)} & \colhead{(deg)} & \colhead{(deg)} & \colhead{ } & \colhead{ } & \colhead{ } & \colhead{ } & \colhead{} & \colhead{ } & \colhead{ }
}
\startdata
  & & & & WC-type & & & & & \\ \hline
 J17112850-3913168 & -39.2214 & 257.8688 & WC8 & 1.782 & 1.221 & 1.239 & 0.445 & 0.958 & 3.209 \\
 J18361633-0705169 & -7.088 & 279.068 & WC9 & 3.071 & 2.051 & 2.094 & 1.011 & -0.128 & -0.433 \\
 J16123747-4637368 & -46.6269 & 243.1561 & WC9 & 1.525 & 1.325 & 1.265 & 0.873 & 0.411 & 0.123\\
 J19202932+1412061 & 14.2017 & 290.1222 & WC-06-d? & 2.358 & 1.352 & 1.126 & 0.667 & 0.723 & 1.324 \\
 J13541345-6150018 & -61.8339 & 208.5561 & WC5-6 & 1.537 & 1.202 & 0.896 & 0.441 & 0.032 & 1.938 \\
 J21500557+5042247 & 50.7069 & 327.5232 & WC5 & 0.41 & 0.711 & 0.238 & 0.375 & 0.754 & 0.726 \\
 J17571686-2523135 & -25.3871 & 269.3203 & WC7: & 2.168 & 1.255 & 0.568 & 0.501 & 1.421 & 1.392 \\
 J10103191-6038423 & -60.6451 & 152.633 & WC5 & 0.19 & 0.569 & 0.312 & 0.331 & 0.769 & 0.689 \\
 J18454987-0259560 & -2.9989 & 281.4578 & WC8 & 0.614 & 1.777 & 1.676 & 0.572 & -0.647 & 2.575 \\
 J18120241-1806554 & -18.1154 & 1,665 & WC8+O8/9III/V & 0.733 & 0.791 & 1.062 & 0.682 & -0.052 & 0.277 \\
 J19122407+0957290 & 9.9581 & 288.1003 & WC & 1.657 & 1.191 & 1.181 & 0.585 & 0.46 & 2.003 \\
 J16352331-4809180 & -48.155 & 248.8472 & WC8 & 2.137 & 1.336 & 0.999 & 0.537 & 0.438 & 1.178 \\
 J18334763-0923077 & -9.3855 & 278.4485 & WC8 & 1.657 & 1.163 & 0.699 & 0.584 & -0.051 & 0.963 \\
 J20115352+3611505 & 36.1974 & 302.973 & WC8+OB: & 0.119 & 0.45 & 0.254 & 0.457 & 0.623 & 0.801 \\
 J14203074-6048221 & -60.8061 & 215.1281 & WC9 & 1.987 & 1.249 & 1.119 & 0.648 & 0.794 & 1.33 \\
 J13520184-6226487 & -62.4469 & 208.0077 & WCE & 2.175 & 1.418 & 0.895 & 0.543 & 0.751 & 2.342 \\
 J17113590-3911075 & -39.1854 & 257.8996 & WC8 & 1.266 & 1.271 & 1.155 & 1.12 & 0.722 & 0.183 \\
 J12300386-6250171 & -62.8381 & 187.5161 & WC7 & 1.11 & 0.962 & 0.551 & 0.449 & 1.064 & 1.765 \\
 J17072379-3919498 & -39.3305 & 256.8492 & WC9 & 1.37 & 0.865 & 0.753 & 0.476 & 0.63 & 1.412 \\
 J14043667-6129165 & -61.4879 & 211.1528 & WC8 & 2.871 & 1.71 & 1.24 & 0.81 & 0.369 & 2.744 \\
 J14565519-5550585 & -55.8496 & 224.23 & WC6.5 & 0.467 & 0.702 & 0.587 & 0.349 & 0.424 & 0.754 \\ \hline
  & & & & WN-type & & & & & \\ \hline
 J18311653-1009250 & -10.1569 & 277.8189 & WN8 & 0.806 & 0.655 & 0.812 & 0.172 & 0.656 & 1.505 \\
 J18420827-0351029 & -3.8508 & 280.5345 & WN8-9h & 1.585 & 0.993 & 0.791 & 0.631 & 0.885 & 3.267 \\
 J19172235+1213133 & 12.2204 & 289.3431 & WN9 & 2.545 & 1.364 & 0.968 & 0.564 & 0.932 & 1.061 \\
 J10255650-5748435 & -57.8121 & 156.4854 & WN5ha/O3+O3Vz((f*)) & 0.608 & 0.367 & 0.461 & 0.158 & 0.863 & 1.740 \\
 J18410086-0426145 & -4.4374 & 280.2536 & WN7 & 0.493 & 0.399 & 0.484 & 0.302 & 0.674 & 0.792 \\
 J15231661-5744198 & -57.7388 & 230.8192 & WN6+O5.5/6 & 0.568 & 0.403 & 0.588 & 0.260 & 0.610 & 0.842 \\
 J18305320-1019370 & -10.3270 & 277.7217 & WN6 & 0.768 & 0.479 & 0.686 & 0.222 & 0.528 & 1.323 \\
 J16285324-4833356 & -48.5599 & 247.2219 & WN5b & 1.851 & 1.036 & 1.003 & 0.366 & 1.796 & 1.887 \\
 J16470761-4549222 & -45.8228 & 251.7817 & WN6h & 2.540 & 0.691 & 1.741 & 1.987 & 3.482 & 1.144 \\
 J16434036-4557576 & -45.9660 & 250.9182 & WN6 & 0.857 & 0.567 & 0.580 & 0.456 & 0.600 & 1.109 \\
 J17142539-3809499 & -38.1639 & 258.6058 & WN6 & 1.526 & 0.846 & 1.196 & 0.374 & 0.913 & 4.092 \\
 J16345746-4704129 & -47.0703 & 248.7394 & WN6 & 2.043 & 1.244 & 1.046 & 0.621 & 0.517 & 1.023 \\
 J12143309-6258509 & -62.9808 & 183.6379 & WN6 & 1.280 & 0.766 & 1.393 & 0.530 & 0.686 & 3.022 \\
 J10534463-5930428 & -59.5119 & 163.4360 & WN4+O8V & 0.210 & 0.274 & 0.332 & 0.278 & 0.797 & 0.978 \\
 J11555211-6245022 & -62.7506 & 178.9671 & WN6 & 0.886 & 0.625 & 0.799 & 0.420 & 0.760 & 0.405 \\
 J11130361-6214183 & -62.2384 & 168.2651 & WN4b & 0.828 & 0.621 & 0.795 & 0.452 & 1.152 & 1.557 \\
 J18492733-0104207 & -1.0724 & 282.3639 & WN7h & 0.875 & 0.595 & 0.661 & 0.314 & 0.165 & 0.761 \\
 J16441069-4524246 & -45.4069 & 251.0446 & WN7 & 1.060 & 0.711 & 0.752 & 0.505 & 0.917 & 2.762 \\
 J17190052-3848513 & -38.8143 & 259.7522 & WN8h & 0.422 & 0.385 & 0.574 & 0.271 & 0.313 & 0.319 \\
 J18255309-1328324 & -13.4757 & 276.4712 & WN6o & 0.796 & 0.559 & 0.777 & 0.357 & 0.817 & 0.961 \\
 J17184971-3357413 & -33.9615 & 259.7072 & WN8/WC9 & 0.469 & 0.512 & 0.493 & 0.358 & 0.699 & 0.622 \\
 J10443211-5750238 & -57.8400 & 161.1338 & WN5 & 0.566 & 0.597 & 0.653 & 0.437 & 0.843 & 0.867 \\
 J12115407-6317037 & -63.2844 & 182.9753 & WN9 & 1.388 & 0.514 & 0.441 & -0.107 & 1.509 & 2.215 \\
 J11023296-5926209 & -59.4391 & 165.6373 & WN5-s & 0.492 & 0.506 & 0.718 & 0.442 & 0.889 & 1.042 \\
 J18250024-1033236 & -10.5566 & 276.2510 & WN7b & 0.974 & 0.625 & 0.927 & 0.372 & 1.017 & 1.344 \\
 J14465358-5919382 & -59.3273 & 221.7233 & WN7-8h & 1.326 & 0.784 & 0.793 & 0.400 & 0.387 & 0.344 \\
 J16482762-4609227 & -46.1563 & 252.1151 & WN5 & 2.290 & 1.262 & 1.281 & 0.558 & 0.638 & 1.611 \\
 J12121681-6246145 & -62.7707 & 183.0701 & WN4-6 & 1.348 & 0.839 & 0.840 & 0.443 & -0.112 & 2.415 \\
\enddata
\end{deluxetable*}
%%%%%%%%%%%%%%%%%%%%%%%%%%%%%%%%%%%%%%%%%%%%%%%%%%%%%%%%%%%%%%%%%%%%%%%%%%%%%%%%%%%%%%%%%
\section{Conclusion}\label{sec:conclusion}
In this study, we performed ML classification of WR stars from a large dataset (6555) of stellar objects of types ranging from MS to AGBs. For this, we developed a highly efficient and robust XGB classifier based on the IR colors and positional coordinates. The models achieve their highest classification efficiency using only 8 features: RA, Dec, J-H, H-$K_{s}$, $K_{s}$-W1, W1-W2, W2-W3, and W3-W4. These features are found to be the most significant in the identification of our objects. The models also perform significantly better than the basic color-color approach of detecting the WR stars. A similar comparison of the XGB classifier with an RF model shows that the former classifies the WR stars with better accuracy than the latter. The models predict that the major contaminants for WR stars are the hot Be-type stars followed by the O-rich AGB sources. 

We also developed a novel WR-subtype classifier using the XGB algorithm capable of distinguishing between WN and WC subtypes with exceptional accuracy. The model performs a much better task in the classification of WC-type stars than the WN class as the color characteristics of the latter overlap with those of the Be-type stars that act as the dominant class of contaminants. The ML model can easily identify the WR-Late type stars, making it more suitable for detecting WR stars in the MW. The model predicts that WNE-type objects are significant contaminants of the WC-class, while WCL-type objects affect the WN-class. Also, the model is capable of identifying the non-classical WNh-type stars. We also applied the object classifier model on an unlabelled dataset (comprised of 6457 stellar sources) and detected 58 new WR candidates. Also, using the WR subtype classifier, we identified the chemical sub-types of 10 of the detected sources. The identified WR candidates are mainly located in the solar neighborhood and within the MW's Local spiral arm that hosts massive star formation regions. In future work, we plan to do a spectroscopic follow-up for our WR candidates. Also, we plan to develop ML classifiers using the same algorithms to identify WR stars across metal-rich galaxies in the local group.  
\section{Acknowledgements}
The authors thank the reviewer for providing constructive comments and suggestions that significantly improved the work. The authors also thank Dr. Lor\'ant O.\ Sjouwerman for his valuable suggestions and discussions that greatly enhanced the work. This research has made use of the CDS SIMBAD database, VizieR catalogue access tool and cross-match service, Strasbourg Astronomical Observatory, France. This publication makes use of data products from the Two Micron All Sky Survey and Wide-field Infrared Survey Explorer which is a joint project of the University of Massachusetts and the Infrared Processing and Analysis Center/California Institute of Technology, funded by the National Aeronautics and Space Administration and the National Science Foundation. S. Kar and R. Das thank the S. N. Bose National Centre for Basic Sciences under the Department of Science and Technology (DST), Govt. of India, for providing the necessary support to conduct research work. Y.P.\ and R.B.\ acknowledge support from the National Aeronautics and Space Administration (NASA) under grant number 80NSSC22K0482 issued through the NNH21ZDA001N Astrophysics Data Analysis Program (ADAP). 

\appendix\section{WR-subtype model misclassification}\label{appendix}
We present the classification shortcomings of the WR sub-type classifier model based on the XGB algorithm (see Sec.\,\ref{subsec:wr_subtype}) in Table\,\ref{tab:Sptype_FPs}. Further, the obtained results are discussed in Sec.\,\ref{sec:discussion}. 
\begin{deluxetable*}{lcccccccccc}
\tablecaption{FPs detected by the Sub-type classifier model (discussed in Sec.\,\ref{subsec:wr_classifier}). \label{tab:Sptype_FPs}}
\tablewidth{0pt}
\tablehead{
\colhead{Source} & \colhead{Dec} & \colhead{RA} & \colhead{Object type} & \colhead{Sub-type} & \colhead{J-H} & \colhead{H-$K_{s}$} & \colhead{$K_{s}$-W1} & \colhead{W1-W2} & \colhead{W2-W3} & \colhead{W3-W4}\\
\colhead{(2MASS)} & \colhead{(deg)} & \colhead{(deg)} & \colhead{} & \colhead{ } & \colhead{ } & \colhead{ } & \colhead{ } & \colhead{} & \colhead{ } & \colhead{ }
}
\startdata
    & & & & WC-type & & & & & & \\ \hline
    J12433281-6306118 & -63.1033 & 190.8867 & Be & Be & -1.328 & 1.06 & 2.328 & 0.424 & 1.219 & 2.01 \\
    J15013011-5916120 & -59.2700 & 225.3755 & WR & WN4b & 2.014 & 1.182 & 1.104 & 0.559 & 0.061 & 0.537 \\
    J15352652-5604123 & -56.070 & 233.860 & WR & WN7 & 1.455 & 0.929 & 0.999 & 0.403 & 0.704 & 0.495 \\
    J18392341-0602158 & -6.0377 & 279.847 & RSG & M0I & 2.511 & 1.344 & 1.393 & 0.276 & 0.869 & 2.139 \\
    J14162737-6117562 & -61.2989 & 214.1140 & WR & WN5b & 2.206 & 1.383 & 1.068 & 0.649 & 0.732 & 0.221 \\
    J19131919+0955289 & 9.9246 & 288.3299 & WR & WN6 & 2.327 & 1.396 & 1.204 & 0.646 & 0.373 & 0.731 \\
    J17063419-3924373 & -39.4104 & 256.6424 & Mira & O-rich & 2.119 & 1.297 & 1.136 & 0.666 & 0.706 & 0.925 \\ \hline
    & & & & WN-type & & & & & & \\ \hline 
    J22195144+5808535 & 58.1482 & 334.9643 & Be & B0Ve & 0.362 & 0.371 & 0.454 & 0.319 & 0.857 & 0.864 \\
    J10412265-6046325 & -60.7757 & 160.3444 & Be & B2 & 0.356 & 0.346 & 0.579 & 0.39 & 1.028 & 0.941 \\
    J16350555-4717135 & -47.2871 & 248.7732 & WR & WC9d? & 1.985 & 1.156 & 1.128 & 0.579 & 0.603 & 1.172 \\
    J18452531-0323011 & -3.3837 & 281.3555 & RSG & M3Ia & 2.078 & 1.024 & 0.982 & 0.38 & 0.672 & 1.672 \\
    J17532843-2446277 & -24.7744 & 268.3685 & Be & B3/5ne & 0.321 & 0.406 & 0.681 & 0.321 & 0.713 & 0.999 \\
    J17452082-2913424 & -29.2285 & 266.3368 & AGB & - & 2.944 & 2.133 & 1.923 & 0.649 & 0.56 & 3.658 \\
    J16002643-5211099 & -52.1861 & 240.1101 & WR & WC7 & 1.331 & 0.906 & 0.812 & 0.508 & 0.605 & 1.073 \\
    J13101207-6239065 & -62.6518 & 197.5503 & WR & WC5 & 0.964 & 0.776 & 0.607 & 0.481 & 1.766 & 3.607 \\
    J16401711-4620098 & -46.3361 & 250.0713 & WR & WC7 & 3.818 & 1.854 & 1.398 & 0.533 & 3.955 & 4.649 \\
    J18400705-0525346 & -5.4263 & 280.0294 & RSG & M4Ib & 1.52 & 0.681 & 0.677 & 0.493 & 1.382 & 1.751 \\
    J17002524-4219003 & -42.3168 & 255.1052 & HMXB & B2e & 0.352 & 0.364 & 0.713 & 0.199 & 0.74 & 1.556 \\
    J17450900-3150158 & -31.8377 & 266.2875 & WR & WC8 & 0.844 & 0.893 & 0.981 & 0.258 & 0.498 & 1.074 \\
    J18332777-1035243 & -10.5901 & 278.3657 & HMXB & B0.5Ve & 0.511 & 0.344 & 0.31 & 0.367 & 0.767 & 1.082 \\
    J21250244+4427063 & 44.4518 & 321.2602 & Be & B1.5V:nnep & 0.226 & 0.263 & 0.474 & 0.365 & 0.68 & 0.827 \\
    J18411070-0451270 & -4.8575 & 280.2946 & WR & WC9 & 0.88 & 0.725 & 0.758 & 0.464 & 1.277 & 1.699 \\
    J18392955-0544222 & -5.7395 & 279.8732 & RSG & K7Ib & 1.325 & 0.528 & 0.583 & -0.292 & 0.223 & 5.325 \\
    J18352720-0704541 & -7.0817 & 278.8634 & AGB & O-rich & 2.707 & 1.328 & 0.959 & 0.503 & 3.056 & 5.149 \\
    J11061873-6114184 & -61.2385 & 166.5781 & WR & WC7 & 0.769 & 0.537 & 0.887 & 0.243 & 0.623 & 0.833 \\
    J20054514+3554030 & 35.9008 & 301.4381 & Be & B1:III/Ve & 0.206 & 0.306 & 0.538 & 0.411 & 0.802 & 0.866 \\
    J17070515-3545323 & -35.7590 & 256.7715 & Be & B1:IIInne & 0.279 & 0.45 & 0.875 & 0.45 & 0.533 & 0.788 \\
    J11061858-6114138 & -61.2372 & 166.5774 & WR & WC7+OB & 0.769 & 0.537 & 0.887 & 0.243 & 0.623 & 0.833 \\
    J16065065-5231556 & -52.5321 & 241.7110 & Mira & O-rich & 2.054 & 1.13 & 1.304 & 0.525 & 0.622 & 0.862 \\
    J18332830-1024087 & -10.4024 & 278.3679 & Be & B0Ve & 0.419 & 0.451 & 0.52 & 0.316 & 0.847 & 1.839 \\
\enddata
\end{deluxetable*}

\newpage
\bibliography{References}{}
\bibliographystyle{aasjournal}

\end{document}